\documentclass[onecolumn,times,astrosymb]{aastex631}

\usepackage{CJK}
\usepackage{amsmath}
\usepackage{xspace}
\usepackage{amssymb}
\usepackage{xcolor}

\newcommand{\ntot}{\ensuremath{n_{\rm total}}\,}
\newcommand{\kms}{\ensuremath{\rm km\,s^{-1}}}
\newcommand{\neff}{\ensuremath{N_{\rm eff}}\,}
\newcommand{\nday}{\ensuremath{N_{\rm days}}}
\newcommand{\rv}{\ensuremath{\overline{V_{\rm r}}}}

\newcommand{\rvi}{\ensuremath{\overline{V_{\rm r}}_{, i}}}
\newcommand{\dvdt}{\ensuremath{\mathrm{d}v / \mathrm{d}t}}
\newcommand{\ppho}{\ensuremath{P_{\rm pho}}}
\newcommand{\Porb}{\ensuremath{P_{\rm orb}}}
\newcommand{\massfunc}{\ensuremath{f(M_2)}}
\newcommand{\msun}{\ensuremath{M_\odot}}
\newcommand{\vsini}{\ensuremath{v\sin i}}
\newcommand{\Gaia}{\textit{Gaia}\xspace}

\begin{document}
\begin{CJK*}{UTF8}{gbsn}

\title{A Sample of Compact Object Candidates in Single-lined Spectroscopic Binaries from LAMOST Medium Resolution Survey}

\shorttitle{Compact object candidates from LAMOST-MRS}
\shortauthors{Liu et al.}

\author[0000-0002-2912-095X]{Hao-Bin Liu}
\affiliation{Department of Astronomy, Xiamen University, Xiamen, Fujian 361005, People's Republic of China}

\author[0000-0003-3137-1851]{Wei-Min Gu}
\affiliation{Department of Astronomy, Xiamen University, Xiamen, Fujian 361005, People's Republic of China}

\author[0000-0002-2419-6875]{Zhi-Xiang Zhang}
\affiliation{Department of Astronomy, Xiamen University, Xiamen, Fujian 361005, People's Republic of China}

\author[0000-0002-5839-6744]{Tuan Yi}
\affiliation{Department of Astronomy, School of Physics, Peking University, Beijing 100871, People's Republic of China}

\author[0000-0002-7420-6744]{jin-zhong Liu}
\affiliation{Xinjiang Observatory, Chinese Academy of Sciences, 150 Science 1–street Urumqi, Xinjiang 830011, People's Republic of China}

\author[0000-0002-0771-2153]{Mouyuan Sun}
\affiliation{Department of Astronomy, Xiamen University, Xiamen, Fujian 361005, People's Republic of China}

\correspondingauthor{Wei-Min Gu}
\email{guwm@xmu.edu.cn}

\begin{abstract}

The stellar spectra from LAMOST Medium Resolution Survey can be used to search for compact objects in binaries. The LAMOST DR10 catalog includes $>$ 980, 000 targets with multiple medium resolution spectra. We select the targets with large or rapid radial velocity variation, and obtained an input-sample of 1822 sources. We use light curves and spectra to identify and exclude eclipsing binaries and double-lined spectroscopic binaries in the input-sample. We finally derive a catalog of 89 candidates with well-folded radial velocity, which are all single-lined spectroscopic binaries, indicating an unseen companion residing in each system. The mass function of each system can be well constrained based on the radial velocity curve. In our sample, 26 sources have mass function higher than 0.1 \msun, among which 18 sources have ellipsoidal type light curves. In our opinion, compact objects are likely existent in all these 26 binaries, which are worth follow-up identification.

\end{abstract}


\section{Introduction}

Compact objects (black holes, neutron stars and white dwarfs) represent the final products of stellar evolution. The search for compact objects provides significant implications for understanding stellar physics and interstellar medium. 
Notably, astrometric observations from \Gaia 
were used to identify compact objects. For instance, \cite{2023MNRAS.518.1057E} and \cite{2023MNRAS.521.4323E} reported two binaries containing stellar black holes (Gaia BH1 and Gaia BH2) from Gaia data release 3, whose companions are respectively a G-type star and a red giant.
Recently, \cite{2024Panuzzo} discovered a black hole of $32.70\pm 0.82$ \msun\, existed in a wide binary system Gaia BH3.

The Doppler spectroscopy have yielded interesting discoveries. \cite{2014Natur.505..378C} claimed a black hole of $3.8 \msun$\, to $6.9 \msun$\, in MWC 656, which was challenged by \cite{2023A&A...677L...9J} who derived a much lower mass of $0.94 \pm 0.34~\msun$\, for the unseen companion. \cite{2019Natur.575..618L} reported that the binary LB-1 consists of a B-type star and a black hole of $68^{+11}_{-13}~\msun$\, albeit this system is also likely to host two luminous stars \citep{2020A&A...639L...6S}. \cite{Saracino2022} reported the detection of a black hole of $11.1^{+2.1}_{-2.4}$ \msun\, in NGC 1850, which was later shown to be a stripped-star binary \citep{2022MNRAS.511L..24E}. In addition, several black hole candidates are also reported in quiescent binary systems \citep{2019Sci...366..637T,2022ApJ...933L..23L,2022arXiv220707675S}. 

Long-term spectroscopic monitoring has also yielded valuable research in dynamically identifying compact binary, containing a neutron star or a white dwarf \citep{2016MNRAS.458.3808R,2017MNRAS.470.1442C,2017MNRAS.472.4193R,2018MNRAS.477.4641R,2021MNRAS.501.1677H,2022arXiv220611270M}. Increasingly, attention has been given to the search of compact objects through time-domain surveys, particularly via the spectroscopic surveys that frequently generate potential candidates. Through comprehensive advantages  in terms of data potential, such time-domain spectroscopic surveys have great predominance and prospects to identify compact object candidates.

The Large Sky Area Multi-Object Fiber Spectroscopic Telescope (LAMOST), a large spectroscopic survey covering the north sky, offers tens of millions of spectra \citep{2012RAA....12.1197C}. The Data Release 10 (DR10) represents a significant milestone in the project, indicating the accumulation of a large volume of observational data and the completion of data processing and analysis up to that release. The DR10 provides multiple spectral observations for individual targets in the Low Resolution Survey (LRS) and the Medium Resolution Survey (MRS). The MRS observation is conducted in about 50\% nights of the LAMOST second stage (since 2018) and have a resolution of 7500 respectively at 5163 Å (blue band) and 6593 Å (red band), which allows for accurate measurement of radial velocities and improved identification of spectral components. Based on the DR10-MRS data, single-lined spectroscopic binary stars can be easily classified and compact objects can be potentially identified in them.

Previous researches on LAMOST data have made significant progresses, with several compact binary candidates discovered, where the unseen companions are neutron stars and high-mass white dwarfs \citep{2019ApJ...872L..20G,2019AJ....158..179Z,2022SCPMA..6529711M,2022ApJ...933..193Z,2022ApJ...940..165Y,2022ApJ...938...78L,2023AJ....165..187Q,2023AJ....165..119Y,2024ApJ...961L..48Z}. With the accumulation of data from LAMOST and ongoing observations, there are more compact object candidates yet to be identified.
Especially in DR10-MRS, which includes a decade of LAMOST spectral data, it provides an excellent opportunity for searching for compact object binary candidates. Moreover, the DR10-MRS data has relatively high spectral resolution and single-exposure data from the same day, which not only meets the precision requirements for radial velocity measurements but also provides supplementary data for the short-period binary star dynamical analysis. We collect the DR10-MRS data as much as possible, and try to search for the compact object candidates in it. The benefit of this search strategy represents on the larger sample coverage and observation data space to obtain more compact objects candidates.

The crux, however, is to overcome the abundance of false data and spurious samples. Previous work relies on designed selecting methods or complex parameter spaces to circumvent most of the false data. In this work, we are shifting focus to radial velocity curves, making everything less cumbersome. We introduce how we generate input-sample relying on raw data from public DR10-MRS catalog and how our pipeline works in Section \ref{pip}. We try to fit the radial velocity curve for every target in input-sample, and then exclude those double-lined spectroscopic binaries and eclipsing binaries. As a result, we obtain a sample with a size of 95 for compact object binary candidate. Our results are displayed in Section \ref{result}. We summarize and discuss our results in Section \ref{conclusion}. Details and techniques are provided in the Appendix.
\section{Candidate Selection}
\label{pip}

We download a complete LAMOST MRS General Catalog, named \textit{dr10\_v1.0\_MRS\_catalogue}, from LAMOST DR10's data access, and remove invalid data in it. In total, we obtain a catalog containing 1134865 sources, including 984719 with repeated observations. 
In this catalog, 130332 sources have more than 3 medium-resolution observational nights and 85062 sources have more than 5 nights. In each observational night, LAMOST conducts commonly 3 medium-resolution exposures for each source. This provides abundant time-domain spectral observation data for a large number of targets, offering a substantial data foundation for dynamic analysis.
In this work, we are dedicated to comprehensively searching for compact object candidates in binaries among sources in DR10. We aim to select candidates from those sources with repeated observations, and the targets should satisfy the following conditions:

\begin{itemize}
    \item  \textit{Requirement}: Radial velocity data should be well-folded.
    \item  \textit{Exclusion}: Consisting of two main-sequence stars.
\label{core}
\end{itemize}

We aim to filter spectroscopic binary through the variation in radial velocity and exclude those consisting 
of two main-sequence stars, including eclipsing binaries and double-lined spectroscopic binaries. As for the first condition, starting from the spectra, we tend to select targets with well-folded radial velocity data, which exhibit relatively high acceptability in terms of their orbital parameters. As for the second condition, we adopted the approach as described in \cite{2022ApJ...940..126F} to exclude: visually inspecting eclipsing binaries by folding photometric data and exclude double-lined spectroscopic binaries referred to the Cross-Correlation Function (CCF) peaks and profiles.

Here, we qualitatively describe our processing workflow as follows. For each target, We collect photometric data and search for its photometric period \ppho.
We consider \ppho\, as a potential orbital period, no matter which specific pattern exhibited in the light curve. Then we search the period of the radial velocity data based on \ppho\, or without a reference period, and fold radial velocities to fit radial velocity curve. We select possible candidates among those targets have smaller fitting residuals. 
After we have selected targets whose radial velocities can be well-folded, we then rule-out those targets with inconsistencies existing in spectroscopy and photometry.

For this investigation, we do not expect to narrow down the input-sample size by individually inspecting. Instead, we attempt to directly fold radial velocities, fit velocity curves, and utilize fitting residuals to select targets.
We select targets with well-folded radial velocities before verifying them for single-lined spectroscopic binaries, which has alleviated several manual inspections of photometric types and spectroscopic data. 
 we reduce artificial involvement with imperative visual inspections positioned as the ultimate step. This strategic approach is devised to contend with the large volume of sources within the DR10 catalog.

\subsection{Input-sample Reduction}


We do the reduction for that targets without spectroscopic binary characteristics, and reduce the catalog size down to the an input-sample. We notice that, in DR10 catalog, there is a high proportion of targets without an obvious radial velocity varying. It is hard to reliably complete periodic analysis for such targets if the observational frequency is not sufficiently abundant. Thus, we prioritize the selection of targets with significant radial velocity variations to enhance the reliability of periodic analysis, especially for those targets with relatively fewer observations.

We obtained the input-sample by screening of the DR10-MRS catalog: We calculate the maximum radial velocity variation for each target by the parameters given from DR10-MRS. In DR10-MRS, each target generally get exposures not less than three times on the same day. 
The catalog provides eight radial velocity measurements and errors, which are also in the following LAMOST MRS Parameter catalog, including \texttt{rv\_b0}, \texttt{rv\_r0}, and so on, corresponding to radial velocities estimated with the B and R band spectra, and others, respectively. The catalog also provides local modified Julian minute (\texttt{lmjm}) and signal-to-noise ratio (\texttt{snr}), etc, for every single exposure. To mitigate the effect caused by measurement errors, we calculate the weighted average radial velocity \rvi\, for the observation day $i$ based on all available exposures:
\begin{equation}
    \rvi = \sum_{j=0}^{n_i} (\texttt{rv\_r0}_{i,j} \cdot \texttt{snr}_{i,j}) \, \bigg/ \,{\sum_{j=0}^{n_i} \texttt{snr}_{i,j}}    
\end{equation}
where $\texttt{rv\_r0}_{i,j} $ and $ \texttt{snr}_{i,j}$ are radial velocity (\texttt{rv\_r0}) and signal-to-noise ratio (\texttt{snr}) of the exposure $j$ on the
observation day $i$.

We exclude ineffective data and exposures of poor quality with $\texttt{snr} \leq 10$. After eliminating ineffective data, we count the number of effective exposures. For each target, we note the number of observation days as \nday, the number of exposures on observation day $i$ as $n_i$ (i = 1,2...\nday) and the all exposure number as \ntot (i.e., $\ntot = \sum_{i} n_i $). 

Typically, a target has not less than three exposures on the same day ($n_i \geq 3$). 
However, after data reduction by \texttt{snr}, there are some observation days that have only one exposure remained (i.e., $n_i = 1$), and this remaining one always keeps worthless measurement so that we are no longer counting the day with only a single exposure into effective observation days \neff. Thus we count effective observation days \neff with enough exposures at a minimum of two exposures ($0 \leq \neff \leq \nday$). 

We apply these two constructed statistics \neff\, and \rvi\ for filtering criteria, to select targets with enough exposures and large radial velocity variations:
\begin{itemize}
    \item Criterion 1: $\neff\, \geqslant \, 5$
    \item Criterion 2: $\Delta \rv \, = \, \overline{V_{\rm r}}_\mathrm{, max} - \, \overline{V_{\rm r}}_\mathrm{, min} >\, 100 \,  \mathrm{\kms}$
\end{itemize}

Furthermore, we notice that some targets still exhibited significant radial velocity variations within a short time ($\sim$ hours), but the lack of observations resulting in not meeting Criterion 2. These targets still hold the potential to include compact objects. Hence, we supplement a set of filtering criteria to encompass short-period targets. We calculate the weighted average rate of radial velocity change \dvdt\, based on the radial velocity (\texttt{rv\_r0}), observation time (\texttt{lmjm}) and signal-to-noise ratio (\texttt{snr}) of all available exposures on the day $i$ and applied the following filter to select such targets:
\begin{equation}
    \frac{\mathrm{d}v}{\mathrm{d}t} \bigg|_i\, = 
    \mathrm{Mean}\left(\frac{\Delta \texttt{rv\_r0}}{\Delta \texttt{lmjm}}\right) \bigg|_i\, = \sum_{j=0}^{n_i} \left(\frac{\texttt{rv\_r0}_{i,j} - \texttt{rv\_r0}_{i,j-1}}{\texttt{lmjm}_{i,j} - \texttt{lmjm}_{i,j-1}} \cdot \,\texttt{snr}_{i,j}\right) \, \bigg/ \,  {\sum_{j=0}^{n_i} \texttt{snr}_{i,j}}
\end{equation}
where $\texttt{rv\_r0}_{i,j} $ and $ \texttt{snr}_{i,j}$ are radial velocity (\texttt{rv\_r0}), observation time (\texttt{lmjm}) and signal-to-noise ratio (\texttt{snr}) of the exposure $j$ on the observation day $i$. We also need to exclude false samples selected due to bad data points, specifically those abrupt \dvdt\, changes deviated significantly from the mean \dvdt. To achieve that, we utilize two statistical parameters, the mean and the variance, to construct criterion:
\begin{equation}
   \mathrm{Var}\left(\frac{\Delta \texttt{rv\_r0}}{\Delta \texttt{lmjm}}\right) < 0.1 \cdot \left(\mathrm{Mean}\left(\frac{\Delta \texttt{rv\_r0}}{\Delta \texttt{lmjm}}\right)\right)^2
\end{equation}
We only consider observation days with more than three exposures ($n_i \geq 3$) to prevent  accidental \dvdt\, error caused by insufficient \texttt{rv\_r0} values. The details of the \dvdt\, and variance filter is described in Appendix \ref{val}. 

Thus we complement these targets selected by two statistic criteria:
\begin{itemize}
    \item Criterion 1': $(\dvdt)_\mathrm{max}\, > \, 60\, \mathrm{km\cdot s^{-1}\cdot h^{-1}} $
    \item Criterion 2': $ \mathrm{Var}(\dvdt) / \mathrm{[Mean}(\dvdt)]^2 < \, 0.1$
\end{itemize}

Adding targets selected by \dvdt\, into input-sample, we obtain the input-sample with the size of 1822. Every target in input-sample have either large enough $\Delta \rv$\, or large enough \dvdt. Then we search the photometric and spectroscopic data to fit radial velocity in the following sections. 

\subsection{Photometric Data}

Photometric data can provide helpful information for the estimate variations and periodic analysis for the close binary. Such systems undergoing orbital motion typically manifest periodic effects in their light curves, such as eclipses and ellipsoidal modulations \citep{1985ApJ...295..143M,1993ApJ...419..344M,2021MNRAS.501.2822G,2021MNRAS.507..104R}.
We make efforts to collect photometric data from several sky surveys, including TESS \footnote{\url{https://tess.mit.edu/science-area/getting-started-with-tess/}}(Transiting Exoplanet Survey Satellite), ZTF\footnote{\url{https://irsa.ipac.caltech.edu/cgi-bin/Gator/nph-scan?projshort=ZTF&mission=irsa}} (Zwicky Transient Facility), and ASAS-SN\footnote{\url{https://asas-sn.osu.edu/variables}} (All-Sky Automated Survey for Supernovae), to facilitate accurate period analysis. For each target in our sample, we collect photometric data as follows:
\begin{itemize}
    \item We use the module \texttt{tesscut} \citep{2019ascl.soft05007B} in package \texttt{LIGHTKURVE} \citep{2018ascl.soft12013L} to download TESS Full-Frame-Images (FFIs) with a cutout-pixel size of 10 from every available sectors in the TESS mission. We utilized the function \texttt{create\_threshold\_mask} to generate target masks and background masks with thresholds set at 15 and 0.001 in FFIs, respectively. We extracted the corresponding fluxes  from each, completing the subtraction of background noise and instrument noise.
    \item We use the package \texttt{ASAS-SN SkyPatrol} \citep{2023arXiv230403791H} to download light curves in ASAS-SN database. We employed the \texttt{cone\_search} function to cross-match targets with the ASAS-SN database, utilizing a cross-matching radius of 2''. We proceed to download the data if there are 20 or more observations of flux.
    \item We download light curves in ZTF DR19 through the ZTF Lightcurve Queries\footnote{\url{https://irsa.ipac.caltech.edu/docs/program_interface/ztf_lightcurve_api.html}} in IRSA (Infrared Science Archive \citep{2018cwla.conf...25T}). We use   to obtain the available g-band\&i-band lightcurves within 5 arcsec of a target position through using the IRSA API.
\end{itemize}

Following the flux correction for data cleaning and time calibration across different time systems, we utilize the Lomb-Scargle algorithm  \citep{1976Ap&SS..39..447L,1981ApJS...45....1S} to search for possible period to fold the light curves.
If the folded light curve from the photometric data exhibited a clean and complete periodic profile, we accept the derived period value. Several samples for periodic light curve are displayed in Appendix \ref{pho}.

Photometric data with a large time span will privide accurate periods, such as those from ASAS-SN and/or ZTF. However, ASAS-SN and ZTF have limited sky coverage for input-sample, thus many targets could not be searched for periods using these two surveys. TESS provides light curve data for nearly every target, but it has a limiting magnitude of only about 15 in TESS band and the time span of each sector is less than 30 days. So there is a risk associated with using TESS period to fold radial velocity, particularly for the target with insufficient spectroscopic observations. To obtain more accurate periods, we attempt to combine data from all available sectors of TESS for each target, enabling the time span is as large as possible to cover the spectroscopic observations. 

\subsection{Radial Velocity Measurement}

We have obtained the photometric period \ppho\, from photometric analysis. We consider it as possible associated orbital period and use it to fold the radial velocity data in order to obtain the radial velocity curve. The DR10-MRS provides radial velocity measurements for both the red and blue arms of each single exposure. We have generated an input-sample used these measurements given by DR10-MRS catalog. 
In fact, there are still numerous imprecise (or incorrect) radial velocity measurements provided in this catalog, possibly due to poor template matching, especially for those spectroscopic binaries with multiple components in observed spectra.

Considering that we do not focus on the barycentric velocity (center-of-mass radial velocity) in this step, we can  improve reliability and measurement accuracy by measuring relative radial velocities.
We pick the spectrum with the highest signal-to-noise ratio \texttt{snr} as the template, and measure the radial velocities of each spectrum from the CCF between them and the template.
In this process, we use the package \texttt{LASPEC} \citep{2021ApJS..256...14Z} to do the reduction on removing cosmic rays and the normalization. We use CCF to measure the relative radial velocities of all spectra for each target in input-sample. As a theoretical template is not necessary, we obtain more accurate radial velocities for a large number of targets with lower time consumption.

Simultaneously, we plot spectra that were shifted based on the radial velocity measurements for final visual inspection. We compare the shifted spectra with template to verify whether the profiles of absorption lines have changed or correctly overlapped after phase calibration. Multiple components in the spectrum could be easily identified, as the absorption line profiles would often get broader (or narrower), and the CCF structures show double peak structures. These phenomena suggest that the system could be a double-lined spectroscopic binary, as there are additional components present in the spectra.
The shifted spectrum, which clearly coincides with the template, indicates measurements are reliable. The radial velocity measurements will ultimately be adopted if they pass the verification through the shifted spectra and CCF profiles for those final candidates. In Appendix \ref{spec_CCF}, we provide examples illustrating the double-peaked structure of the CCF for double-lined spectroscopic binary spectra.

It is worth noting that the Radial Velocity Zero-Point (RVZP) of MRS may vary over time and with different fibers \citep{Zhang_2021}. The RVZP correction usually introduces an average uncertainty of around 5 \kms\, to radial velocity measurements \citep{Zhang_2021}. Our selected sources have a typical radial velocity semi-amplitude of 100 \kms, which is significantly larger than 5 \kms. Thus we do not take RVZP into account in radial velocity measurements. 


\subsection{Radial Velocity Curve Fits}
We fit folded data for the radial velocity curve based on a sinusoidal function:

\begin{equation}
    V_{\mathrm r}(t) = V_{0} + K \sin{\left(\frac{2 \pi t}{\Porb} + \phi_0\right)}
\end{equation}
where \Porb\, is the orbital period, $K$ is the semi-amplitude of the radial velocity curve, $V_0$ is the offset radial velocity generated by the template selection and $\phi_0$ is initial phase of the folding starting point.
This model assumes that the orbit is circular, which is justified,
since ellipsoidal binaries undergo tidal interactions that both circularize the orbit and synchronize the rotational and orbital period \citep{Zahn1977, Lurie_2017}. 
Ideally, the time span of photometric data acquisition should cover the spectroscopic observations. In LAMOST DR10-MRS, the time span of individual target spectroscopic observations could be relatively large ($\thicksim 10^3$ days). Therefore, the best time span for photometric data should also be greater than this length.

The data of ZTF and ASAS-SN are best-suited for searching a precise period because they have thousands of days of observational data. If a target can obtain its photometric period \ppho\, from ZTF or ASAS-SN, this period will be directly regarded as a preferred period for folding the radial velocity. However, if the target is relatively faint resulting in bad data in the ground-based photometric surveys, and the photometric period can only be obtained from TESS, then this period cannot be directly used to fold the radial velocity. In this case, we search for orbit period \Porb\, using radial velocity data within a small neighborhood of the reference period in order to obtain a reliable radial velocity curve. Several samples for radial velocity curve are displayed in Appendix \ref{rvcurve}. We show radial velocity curves for those target with at least 0.1 \msun mass function in Appendix \ref{rv_mass}.

Still, for some targets, there is no essential relation between photometry and orbit, whose \ppho\, and \Porb\, arise from different physics. For targets that cannot be folded by the photometric period \ppho, we directly utilize the radial velocity data for searching period. We employ the \texttt{Lomb-Scargle}  algorithm to search for potential periodic signals within the range of 0.1 days to 100 days. We fold the radial velocities with the period corresponding to the highest Lomb-Scargle power to obtain the radial velocity curve.

In cases where there is no photometric period available or a photometric period obtained by TESS, the quality of the radial velocity curve depends more on the number of radial velocity observations. If there are too few radial velocity data points, the fitting of the radial velocity curve may have multiple solutions, leading to overfitting or under-fitting. Hence, we still select the candidates with more radial velocity observations in a higher priority. 

\section{Results}
\label{result}
The above procedure produces possible radial velocity curves for every target in our input-sample. After period search and folding, over 1500 targets did not show well-converged radial velocity curve profiles, meaning them lack periodic variations in their radial velocity observations, possibly due to insufficient exposures or incorrect adopted period. We conducted a manual review of approximately 300 targets with complete radial velocity curves, excluding those exhibiting variability attributed to various types of eclipses (As shown in Figure \ref{spec} in the Appendix \ref{spec_CCF}) and those with double-lined spectroscopic spectra. We do the spectroscopic exclusion by proofreading shifted spectra and double-peaked structure in the CCF. We have provided an explanation of the types of CCF structure in the Appendix \ref{spec_CCF}.

We select the candidates from the input-sample, and through manual examination of both photometric and spectroscopic data, we eliminate potential spectroscopic binaries, leaving us with 89 candidates. These targets all possess a complete folded velocity curve with a sufficient number of radial velocity points. 64 targets also have light curves that synchronize with the orbit phase, indicating orbital modulation in the photometric variation. For other targets, either no light curve is available or the light curve does not show clearly converge at some period. For these targets, the orbital period is primarily determined through period searching in the radial velocity data.

Fitting the radial velocity curves yielded orbital parameters for the targets, including the radial velocity semi-amplitude $K$ and orbital period \Porb. Then, the mass of the invisible object can be constrained by the mass function, expressed as \citep{2006ARA&A..44...49R}:
\begin{equation} \label{eq:2}
    f(M_2) = \frac{M_2^3 \sin^3(i)}{(M_1 + M_2)^2} = \frac{K_1^3 \Porb}{2\pi G},
\end{equation}
where $M_1$ is the mass of the optically visible star; $M_2$ is the mass of the unseen companion; $K_1 $ is the semi-amplitude of the radial velocity; $\Porb $ is the orbital period; $i$ is the inclination, and $G$ is the gravitational constant. We calculate the mass function for our sample using parameters obtained from the radial velocity curve fitting. The distribution of all targets in the mass function space is shown in Figure \ref{mass_function}, and listed in Appendix \ref{full_tab}.

\begin{figure}[t!]
\centering
\includegraphics[scale=0.55]{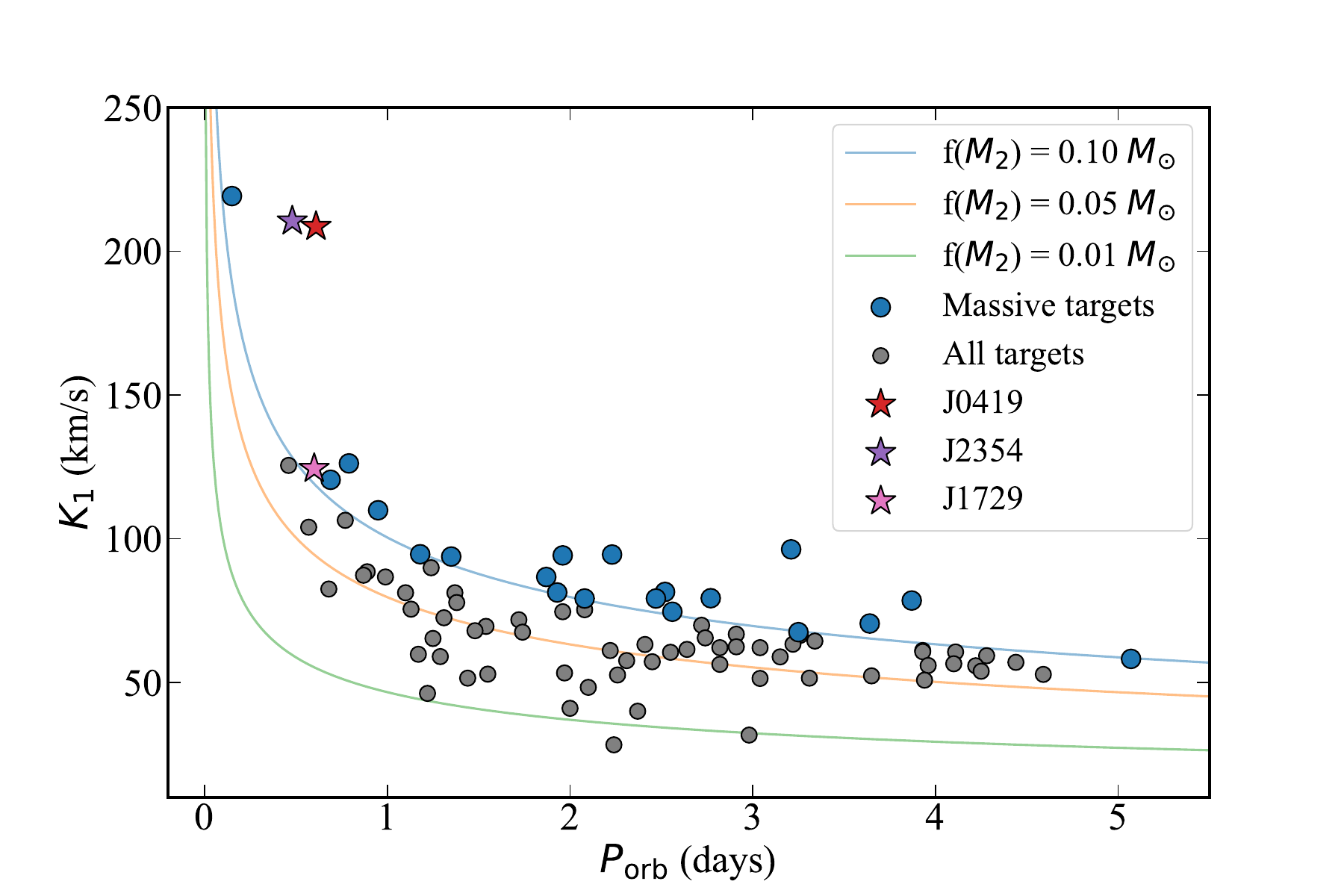}
\caption{Radial velocity semi-amplitude versus orbital period. All selected 89 targets' mass functions are calculated using the parameters shown in this figure. The blue circles denote 26 targets whose mass functions are greater than 0.1 \msun, while the gray circles represent the other candidates and colored asterisks represent three identified candidates. The three colored reference lines, from top to bottom, respectively represent cases where the mass function equals 0.10 \msun, 0.05 \msun\, and 0.01 \msun.
\label{mass_function}}
\end{figure}

In the Figure \ref{mass_function}, we present the distributions of radial velocity semi-amplitudes and orbital periods for all 89 candidates in our sample. The orbital periods of our candidates range from 0.1 days to approximately 10 days, and the radial velocity semi-amplitudes span from about 30 \kms to approximately 200 \kms. Nearly all of the sources fall above the green reference curve, indicating their mass functions are greater than 0.01 \msun. There are 69 targets reside in the region above the orange reference curve which corresponds to a mass function greater than 0.05 \msun. Among these targets, targets No.1-26 in the sample have mass functions exceeding 0.1\,\msun.

In our result, there are three published identifications of inclusive targets (J0419, J2354 and J1729), which are marked with colored asterisks in figures. Target No.1 (J0419) contains an extremely low mass pre-white dwarf  and a compact object, with $M_1 = 0.176 \pm 0.014 \,\msun $ and $ M_2 = 1.09 \pm 0.05 \,\msun$ \citep{2022ApJ...933..193Z}. Target No.2 (J2354) contains an K-type star and a neutron star candidate, with $M_1 = 0.70 \pm 0.05 \,\msun $ and $ M_2 = 1.26 \pm 0.03  \,\msun$ \citep{2023SCPMA..6629512Z}. Target No.18 (J1729) contains an K-type star and a white dwarf, with $M_1 = 0.81 \pm 0.07 \,\msun $ and  $M_2 \geq 0.63  \,\msun$ \citep{Zheng_2022}. 
The remaining targets currently await individual analysis.

In addition, compared to the three identified candidates, the other candidates in our sample can exhibit relatively longer orbital periods. Leveraging tens of observations provided by LAMOST and employing period analysis, we also obtained targets with relatively minor radial velocity variations. As shown in Table \ref{tab_lmf}, except for Target No.18 (J1729), which only has one day of effective observational data, the rest of the targets have multiple days of LAMOST observations. Half of the targets have more than ten days of spectroscopic observations, with dozens of radial velocities. We examined the spectra of the candidates to ensure accurate radial velocity measurements. Well-folded radial velocity data provided reliable orbital parameters. We have ruled out double-lined spectroscopic binaries among the candidates in our sample, and it is worth to further identify our candidates, especially for candidates resembling the green asterisk marker (J1729) in the Figure \ref{mass_function}.

We collect the parallax and the stellar parameters including effective temperature $T_{\rm{eff}}$, surface gravity log$g$, metallicity [Fe/H], and radius from \Gaia DR3 \citep{2022arXiv220800211G}. We use the stellar evolution models to evaluate the mass of visible stars by utilizing the python package \texttt{isochrones}\footnote{\url{https://isochrones.readthedocs.io/en/latest/}}. We use the \Gaia DR3 parameters and photometry as inputs to derive isochrone interpolated mass with MESA Isochrones \& Stellar Tracks isochrones \citep[MIST;][]{2016ApJS..222....8D}. The isochrone mass of visible star is shown in Table \ref{tab_mass}. Combined with the mass function equation, the lower limit of the invisible object's mass $M^{\rm{min}}_2$ is calculated under the assumption that $i=90^{\circ}$. We find that the lower limit of the invisible object's mass $M^{\rm{min}}_2$ is above or close to half of the visible stellar mass $M_1$ for the first thirty or so targets in our sample. A normal main-sequence star at a such mass would contribute non-negligible optical flux and manifest double-lined spectroscopic features, thus these single-lined spectroscopic binaries could conceal a compact object. The binary mass for target No.1-26 is presented in Figure \ref{mass} and listed in Table \ref{tab_mass}. Parameters for all candidates is presented in in Appendix \ref{mass_all} and Appendix \ref{full_tab}.

\begin{figure}[t!]
\centering
\includegraphics[scale=0.75]{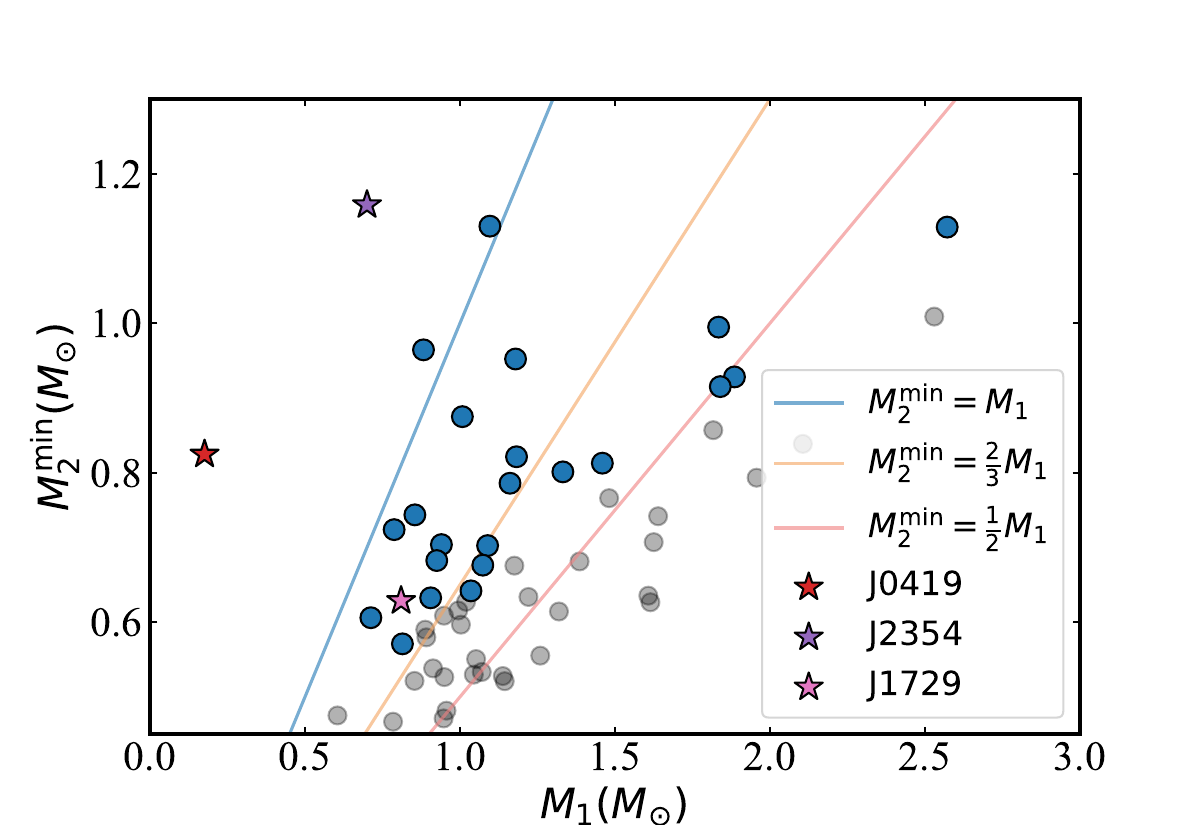}
\caption{The minimum mass of the unseen companion at the maximum inclination ($i = 90^{\circ}$). Three solid lines represent the cases $M_2^\mathrm{min} = M_1$, $M_2^\mathrm{min} = 2 M_1/3$ and $M_2^\mathrm{min} = M_1/2$, respectively. The blue circles in the graph indicate 21 massive targets marked by the same marker in Figure \ref{mass_function}, while the gray circles represent the other candidates and colored asterisks represent three identified candidates.
Most of targets marked by blue circles have the similar lower limit of mass ratio ($M_2^\mathrm{min} / M_1$) to the green asterisk's (J1729), whereas their unseen companion are more massive than it.
\label{mass}}
\end{figure}

\section{Conclusions and Discussion}
\label{conclusion}
In this paper, we have constructed a catalog of 89 compact object candidates in single-lined spectroscopic binaries using the LAMOST DR10-MRS observations.
All candidates have well-folded radial velocity data. Overall, the results are very encouraging: a large number of candidate compact objects are discovered in a single pipeline execution, among which our existing identifications are also included (i.e., target No.1 J0419, target No.2 J2354 and target No.18 J1729). For 64 out of 89 objects, their phase folded light curves are consistent with the radial velocity variations; hence, their orbital parameters (e.g., orbital period and semi-amplitude) are more reliable. Our main results are as follows:
\begin{enumerate}
    \item \textit{Initial target selection}: We utilize data from the DR10-MRS catalog to establish selection criteria to pre-select targets, particularly introducing variability rates \dvdt\, to selective focus on the shorter-period targets. 
    By imposing constraints on the variance of \dvdt\, we additionally obtain a larger number of targets with reliable radial velocity measurements, expanding sample pool for subsequent verification.
    Combining criteria on radial velocity variation, we create an initial sample of 1822 targets, all of which have the potential to possess large mass function because of large or rapid radial velocity variations.
    \item \textit{Compact object candidates}: We ultimately select 89 single-lined spectroscopic binaries with well-folded radial velocity data, which probably contain compact object candidates. The mass function of 26 candidates are larger than 0.1 \msun. Among them, 18 targets also exhibit ellipsoidal type light curves. These candidates fulfill the criteria wherein the minimum limit of the unseen object's mass $M^{\rm{min}}_2$ is above or approximately half of the half of the visible stellar mass $M_1$. This estimation strongly supports the presence of compact objects within these candidates.
    \item \textit{Optimization of selection procedures}： We directly fit radial velocity curves for 1822 targets, selecting those with well-converged curves. We deferred manual inspection (for both photometry and spectroscopy) until after the source selection stage, which saved us from manually inspecting over a thousand targets and greatly reduced the time resource waste caused by erroneous measurements both in photometric and spectroscopic. In order to achieve this goal, we use the workflow described in Section 2 for handling photometric and spectral data, generating reliable radial velocity curves for potential targets.
\end{enumerate}

To further constrain the mass determination of the unseen object, precise inclination measurement is imperative. Ellipsoidal modulation has been detected in part of our candidates, allowing for inclination analysis through fitting light curves. However, a majority of reported candidates lack observable ellipsoidal light curves because the insufficiently close proximity of the binary stars results in tidal deformation not obvious. Besides, there are challenges in accurately constraining the rotational velocity of the visible stars, limiting our ability to determine inclination by measuring \vsini. While this paper focused on compiling the compact object binary catalog, it is important to highlight these potential researches for the sampled candidates, which will be
pursued in the future work.

\section*{acknowledgments}
We thank the anonymous referee for their constructive suggestions to improve the paper.
This work was supported by the National Key R\&D Program of China under grants 2023YFA1607901 and 2021YFA1600401, the National Natural Science Foundation of China under grants 11925301, 12033006, 12221003, 12263003 and 12322303, the Natural Science Foundation of Fujian Province of China under grants 2022J06002, and the fellowship of China National Postdoctoral Program for Innovation Talents under grant BX20230020. J.Z.L was supported by the Tianshan Talent Training Program  through the grant 2023TSYCCX0101.
This paper uses the data from the LAMOST survey.
Guoshoujing Telescope (the Large Sky Area Multi-Object Fiber Spectroscopic Telescope LAMOST) is a National Major Scientific Project built by the Chinese Academy of Sciences. 
Funding for the project has been provided by the National Development and Reform Commission. LAMOST is operated and managed by the National Astronomical Observatories, Chinese Academy of Sciences.
This paper includes data collected by the TESS mission obtained from the Mikulski Archive for Space Telescopes (MAST) at the Space Telescope Science Institute. The specific observations can be accessed via \cite{https://doi.org/10.17909/fwdt-2x66}. 
ZTF is a public-private partnership, with equal support from the ZTF Partnership and from the U.S. National Science Foundation through the Mid-Scale Innovations Program (MSIP).

 \software{
    Astropy \citep{2013A&A...558A..33A,2018AJ....156..123A,2022ApJ...935..167A},
    Lightkurve \citep{2018ascl.soft12013L},
    Matplotlib \citep{Hunter:2007},
    NumPy \citep{harris2020array},
    Pandas \citep{mckinney-proc-scipy-2010},
    ASAS-SN sky patrol \citep{2023arXiv230403791H},
    Laspec \citep{2021ApJS..256...14Z},
    Isochrones \citep{2015ascl.soft03010M}
}

\bibliography{main}{}
\bibliographystyle{aasjournal}

\newpage
  \appendix
  \section{Observation conclusions for 1 - 26 targets}

  \startlongtable
  \begin{deluxetable*}{ccccccccc}
    \tablecaption{Mass functions for 1 - 26 targets}
    \tablewidth{0.99\textwidth}
    \tabletypesize{\scriptsize}
    \tablehead{
      \colhead{Number} & \colhead{Designation} & \colhead{Source ID}      & \colhead{\neff}        & \colhead{$\Delta \rvi$} & \colhead{\Porb} & \colhead{LC} & \colhead{$K$} & \colhead{\massfunc} \\
      \colhead{}          & \colhead{}            & \colhead{}    & \colhead{(1)}       & \colhead{(2)}       & \colhead{(3)}           & \colhead{(4)}     & \colhead{(5)} & \colhead{(6)}  \\
      \colhead{}          & \colhead{}            & \colhead{}    & \colhead{(days)} & \colhead{(km s$^{-1}$)} & \colhead{(days)} & \colhead{} & \colhead{(km s$^{-1}$)}     & \colhead{(\msun)}             
    }
    \startdata
    1     & J041920.07+072545.4 & 410015 & 31    & 418.7 & 0.61  & ELL   & 208.6 & 0.560 \\
    2     & J235456.76+335625.7 & 1129096 & 5     & 426.5 & 0.48  & ELL   & 210.5 & 0.455 \\
    3     & J134417.15+520307.2 & 818528 & 5     & 130.55 & 3.21  &       & 96.3  & 0.291 \\
    4     & J102740.93+384018.9 & 697652 & 11    & 126.44 & 9.97  & ELL   & 63.9  & 0.263 \\
    5     & J122924.56+473128.5 & 771613 & 7     & 185.76 & 2.23  & ELL   & 94.5  & 0.190 \\
    6     & J060826.18+211334.1 & 400747 & 6     & 141.74 & 3.87  &       & 78.5  & 0.189 \\
    7     & J122302.07+482910.8 & 767594 & 10    & 189.98 & 1.96  & ELL   & 94.2  & 0.166 \\
    8     & J060030.98+290855.0 & 387387 & 11    & 447.04 & 0.15  & ELL   & 219.2 & 0.162 \\
    9     & J085304.59+132032.3 & 616838 & 17    & 249.3 & 0.79  & ELL   & 126.2 & 0.161 \\
    10    & J060649.24+213233.3 & 397914 & 12    & 157.38 & 2.77  & ELL   & 79.3  & 0.140 \\
    11    & J060907.13+222550.6 & 401948 & 11    & 128.91 & 2.52  &       & 81.5  & 0.138 \\
    12    & J092323.52+421819.5 & 651359 & 15    & 134.79 & 3.64  & ELL   & 70.5  & 0.129 \\
    13    & J230854.08+355132.4 & 1095399 & 9     & 207.93 & 0.95  & ELL   & 109.9 & 0.128 \\
    14    & J090332.95+425639.8 & 630705 & 35    & 210.12 & 8.41  & ELL   & 53.2  & 0.128 \\
    15    & J044826.19+223906.2 & 296802 & 7     & 154.56 & 2.47  & ELL   & 79.2  & 0.124 \\
    16    & J063350.35+220056.0 & 447835 & 7     & 235.13 & 0.69  &       & 120.5 & 0.123 \\
    17    & J155751.89+435021.4 & 884025 & 6     & 175.58 & 1.87  &       & 86.7  & 0.123 \\
    18    & J172900.17+652952.8 & 905386 & 1     & 94.8  & 0.6   & ELL   & 93.9  & 0.123 \\
    19    & J075855.51+384540.7 & 550320 & 15    & 177.7 & 1.35  & ELL   & 93.8  & 0.113 \\
    20    & J103613.90+071115.6 & 703271 & 10    & 113.53 & 5.68  &       & 57.3  & 0.108 \\
    21    & J061921.71+224432.3 & 422288 & 10    & 163.02 & 2.56  & ELL   & 74.6  & 0.107 \\
    22    & J061125.78+115043.5 & 406053 & 6     & 114.71 & 1.93  &       & 81.3  & 0.105 \\
    23    & J011726.82+452013.2 & 64557 & 7     & 163.61 & 2.08  &       & 79.2  & 0.104 \\
    24    & J060537.69+222546.9 & 395839 & 13    & 136.43 & 3.25  & ELL   & 67.5  & 0.101 \\
    25    & J202309.48+401916.5 & 1010372 & 7     & 134.48 & 1.18  & ELL   & 94.6  & 0.101 \\
    26    & J022924.20+581721.3 & 135083 & 6     & 109.53 & 5.07  & ELL   & 58.2  & 0.101 
    \enddata
    \tablecomments{Column(1) The effective observation days in LAMOST DR10-MRS. Column(2) The maximum difference for effective mean radial velocity \rvi in LAMOST DR10-MRS. Column(3) The orbital period obtained from the radial velocity curve fitting. Column(4) The types of light curve, "Ell" stands for ellipsoidal modulation. Column(5) The semi-amplitude of the radial velocity obtained from the radial velocity curve fitting. Column(6)The mass function of the unseen companion calculated by Column(3) and Column(5).
\label{tab_smf}}
  \end{deluxetable*}
  
\clearpage
  \startlongtable
  \begin{deluxetable*}{cccccccccc}
    \tablecaption{Mass measurements  for 1 - 26 targets}
    \tabletypesize{\scriptsize}
    \tablehead{
      \colhead{Number} & \colhead{Designation} & \colhead{Source ID}      & \colhead{$M_1$}        & \colhead{$M_2^{90}$} & \colhead{$M_2^{80}$} & \colhead{$M_2^{70}$}  & \colhead{$M_2^{65}$} & \colhead{$M_2^{60}$} &\\
      \colhead{}          & \colhead{}            & \colhead{}    & \colhead{(1)}       & \colhead{(2)}       & \colhead{(3)}           & \colhead{(4)}     & \colhead{(5)} & \colhead{(6)}  \\
      \colhead{}          & \colhead{}            & \colhead{}    & \colhead{(\msun)} & \colhead{(\msun)} & \colhead{(\msun)} & \colhead{(\msun)}     & \colhead{(\msun) }   & \colhead{(\msun) }            
    }
    \startdata          
    1     & J041920.07+072545.4 & 410015 &       &       &       &       &       &  \\
    2     & J235456.76+335625.7 & 1129096 &       &       &       &       &       &  \\
    3     & J134417.15+520307.2 & 818528 & 1.10  & 1.13  & 1.16  & 1.24  & 1.31  & 1.41  \\
    4     & J102740.93+384018.9 & 697652 & 0.88  & 0.96  & 0.99  & 1.06  & 1.12  & 1.21  \\
    5     & J122924.56+473128.5 & 771613 & 1.18  & 0.95  & 0.97  & 1.04  & 1.10  & 1.17  \\
    6     & J060826.18+211334.1 & 400747 & 1.01  & 0.87  & 0.89  & 0.96  & 1.01  & 1.08  \\
    7     & J122302.07+482910.8 & 767594 & 0.79  & 0.72  & 0.74  & 0.79  & 0.84  & 0.90  \\
    8     & J060030.98+290855.0 & 387387 &       &       &       &       &       &  \\
    9     & J085304.59+132032.3 & 616838 & 0.85  & 0.74  & 0.76  & 0.81  & 0.86  & 0.92  \\
    10    & J060649.24+213233.3 & 397914 &       &       &       &       &       &  \\
    11    & J060907.13+222550.6 & 401948 & 1.18  & 0.82  & 0.84  & 0.90  & 0.94  & 1.01  \\
    12    & J092323.52+421819.5 & 651359 & 0.94  & 0.70  & 0.72  & 0.77  & 0.81  & 0.86  \\
    13    & J090332.95+425639.8 & 630705 & 1.16  & 0.79  & 0.80  & 0.86  & 0.90  & 0.96  \\
    14    & J230854.08+355132.4 & 1095399 & 0.71  & 0.61  & 0.62  & 0.66  & 0.70  & 0.75  \\
    15    & J044826.19+223906.2 & 296802 &       &       &       &       &       &  \\
    16    & J063350.35+220056.0 & 447835 & 1.83  & 0.99  & 1.01  & 1.08  & 1.13  & 1.20  \\
    17    & J155751.89+435021.4 & 884025 & 0.93  & 0.68  & 0.70  & 0.74  & 0.78  & 0.84  \\
    18    & J172900.17+652952.8 & 905386 &       &       &       &       &       &  \\
    19    & J075855.51+384540.7 & 550320 & 1.33  & 0.80  & 0.82  & 0.87  & 0.92  & 0.97  \\
    20    & J103613.90+071115.6 & 703271 & 1.09  & 0.70  & 0.72  & 0.76  & 0.80  & 0.86  \\
    21    & J061921.71+224432.3 & 422288 & 0.91  & 0.63  & 0.65  & 0.69  & 0.73  & 0.77  \\
    22    & J061125.78+115043.5 & 406053 & 2.57  & 1.13  & 1.15  & 1.22  & 1.28  & 1.36  \\
    23    & J011726.82+452013.2 & 64557 & 1.46  & 0.81  & 0.83  & 0.88  & 0.93  & 0.99  \\
    24    & J022924.20+581721.3 & 135083 & 1.89  & 0.93  & 0.95  & 1.01  & 1.05  & 1.12  \\
    25    & J202309.48+401916.5 & 1010372 & 1.84  & 0.92  & 0.93  & 0.99  & 1.04  & 1.10  \\
    26    & J060537.69+222546.9 & 395839 & 1.07  & 0.68  & 0.69  & 0.74  & 0.77  & 0.82  \\
    \enddata
    \tablecomments{Column(1) The mass of visible star from Gaia. Column(2-6) The mass of unseen companion estimated by Column(1) and inclination of 90°, 80°, 70°, 65° and 60° respectively.
    \label{tab_mass}}
  \end{deluxetable*}

  \clearpage
  \section{Criterion variance}
  Observations conducted within the same day, with intervals much shorter than the orbital period, should showcase uniform variations in radial velocity. However, during our screening process, we noticed abrupt changes in the radial velocity data for most targets across multiple observations on the same day. To address this issue, we devised a statistical metric based on the rate of change in radial velocity to filter out potentially erroneous measurements. For ideal target, the variance in the rate of change of radial velocity for a pristine sample should be sufficiently low. Hence, we employed the variance in radial velocity change rate as a filtering criterion. We presented a case with ample observation instances to validate the effectiveness of this approach.
  
  \label{val}
\begin{figure}[h]
\centering
\includegraphics[scale=0.55]{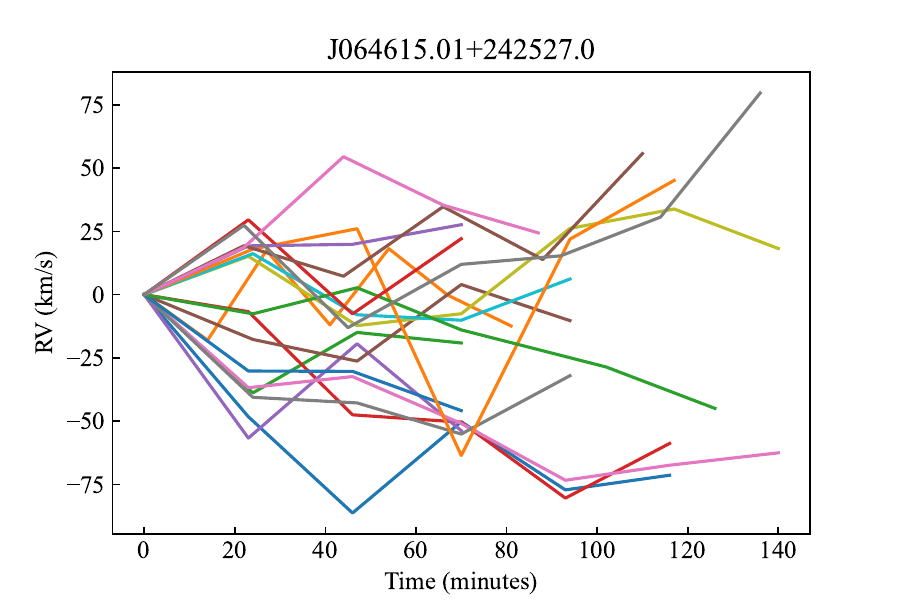}
\includegraphics[scale=0.55]{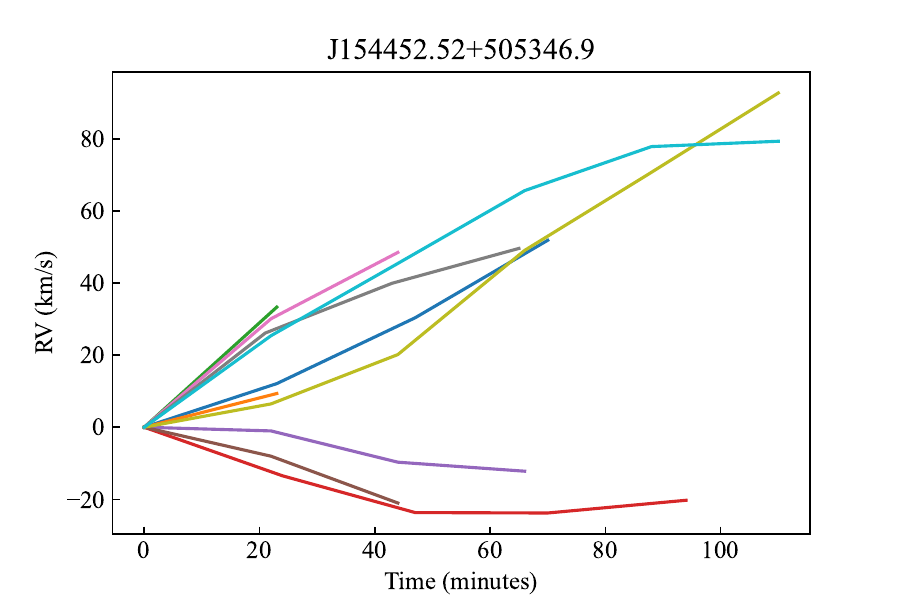}
\caption{A comparative figure. The left panel displays targets obtained without variance restrictions, while the right panel shows targets acquired through variance limitations. Each colored line represents observations on different days. It is evident that the addition of variance-based filtering significantly enhances the smoothness of radial velocity for the selected targets. The left panel is not an extreme counterexample, it demonstrates criterion effectiveness even for targets with a higher number of observations.
\label{Obs_fig}}
\end{figure}

  \section{Radial velocity curves}
  Here we show two targets, for example, amoung our sample with their radial velocity curves and light curves, which are J060649.24+213233.3 (left sub-figure) and J063552.55+184233.7 (righ sub-figure). 
\label{rvcurve}
\begin{figure}
\centering
\includegraphics[scale=0.36]{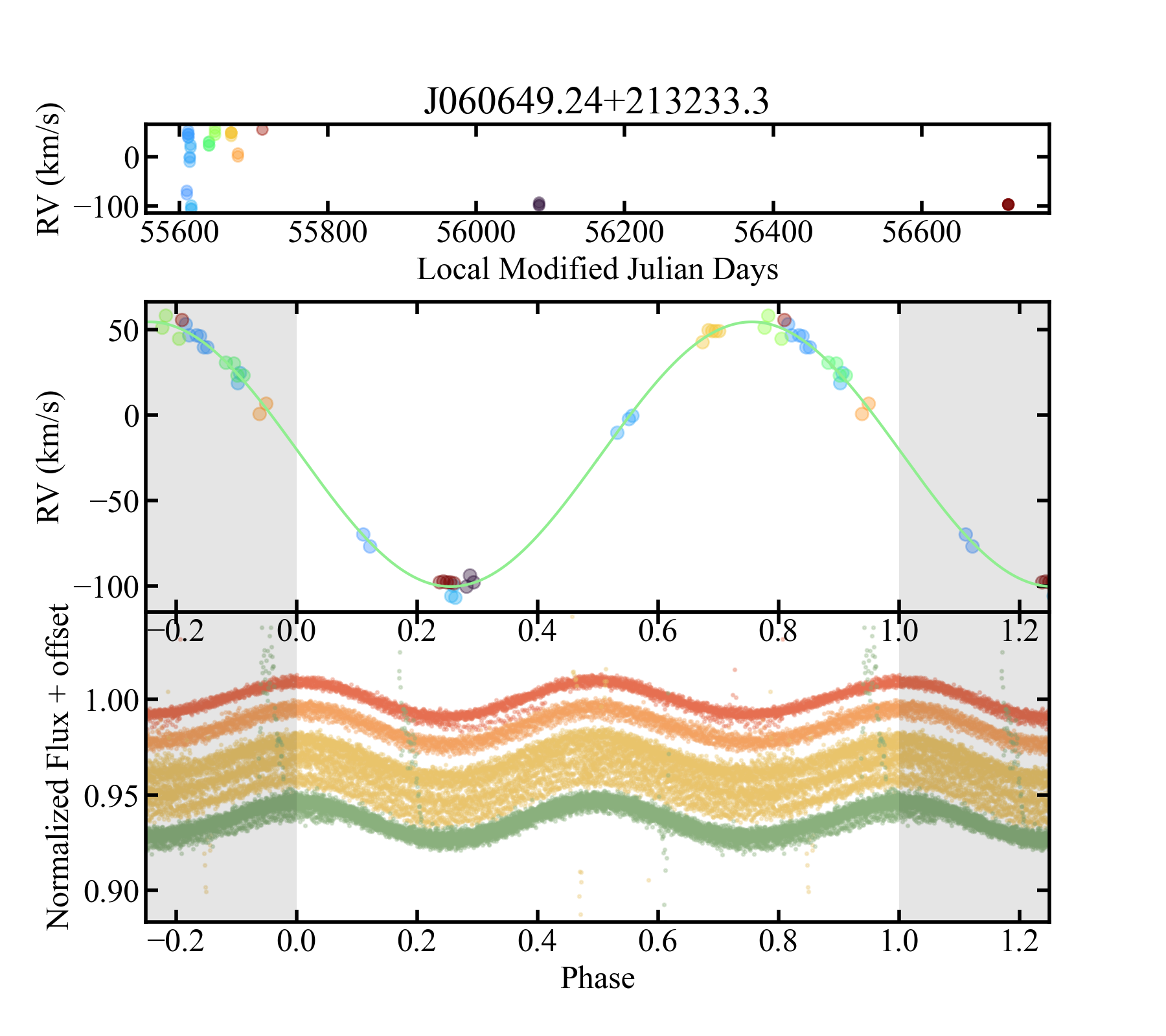}
\includegraphics[scale=0.36]{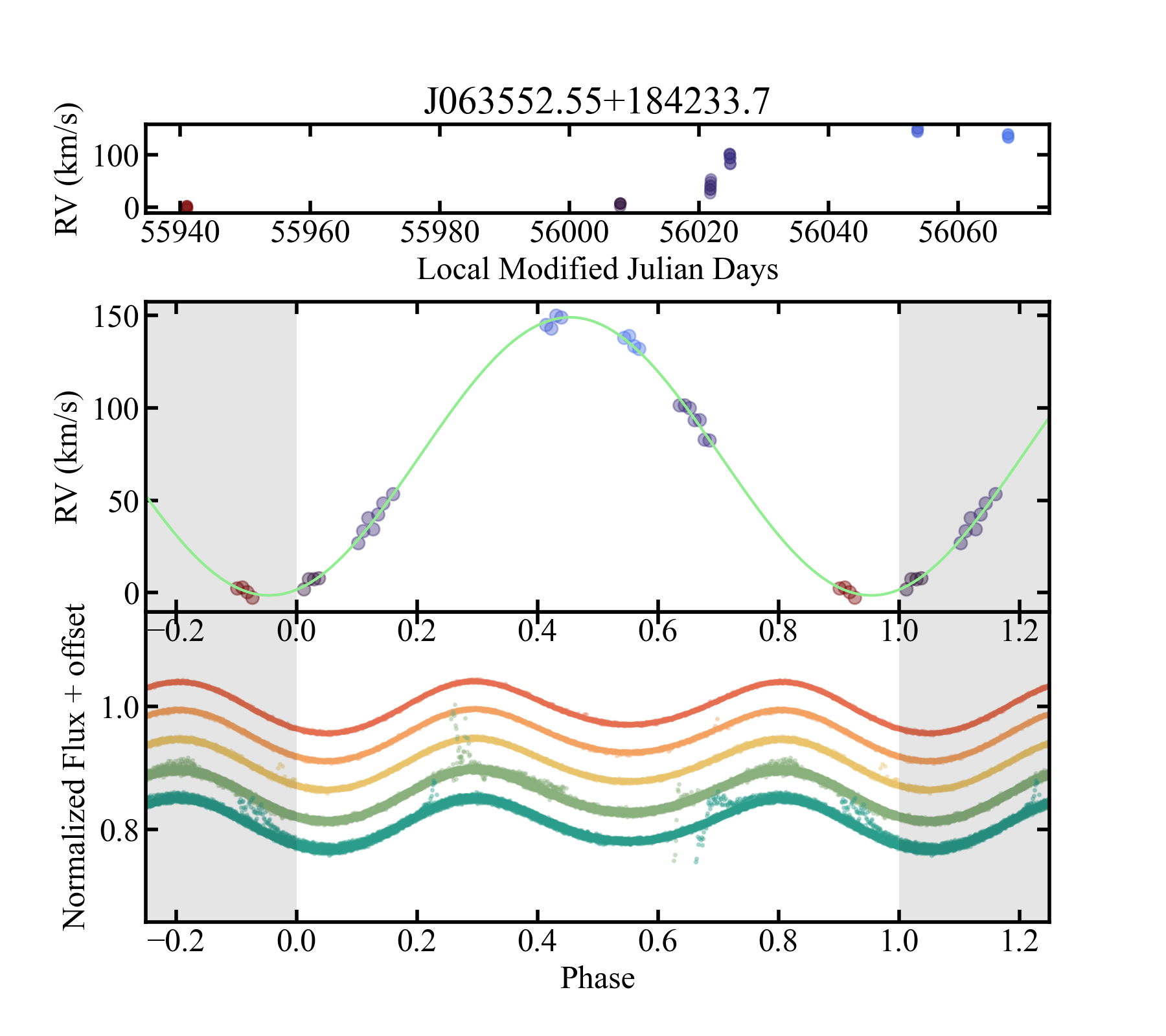}
\caption{For each target, the top panel represents the radial velocity measured by LAMOST DR10-MRS spectroscopy using CCF. The middle one shows the folded radial velocity curve by searched period. The bottom curve corresponds to the folded TESS light curve with the same period. Both photometric data and radial velocities are clearly folded by the same period.
\label{rv_fig}}
\end{figure}

  \section{Light curves}
  \label{pho}
Here we show some folded TESS light curves. we collect photometric data from all availble sectors of TESS, shown in different colors and flux offsets in each figure. 
\begin{figure}[h]
\centering
\includegraphics[scale=0.22]{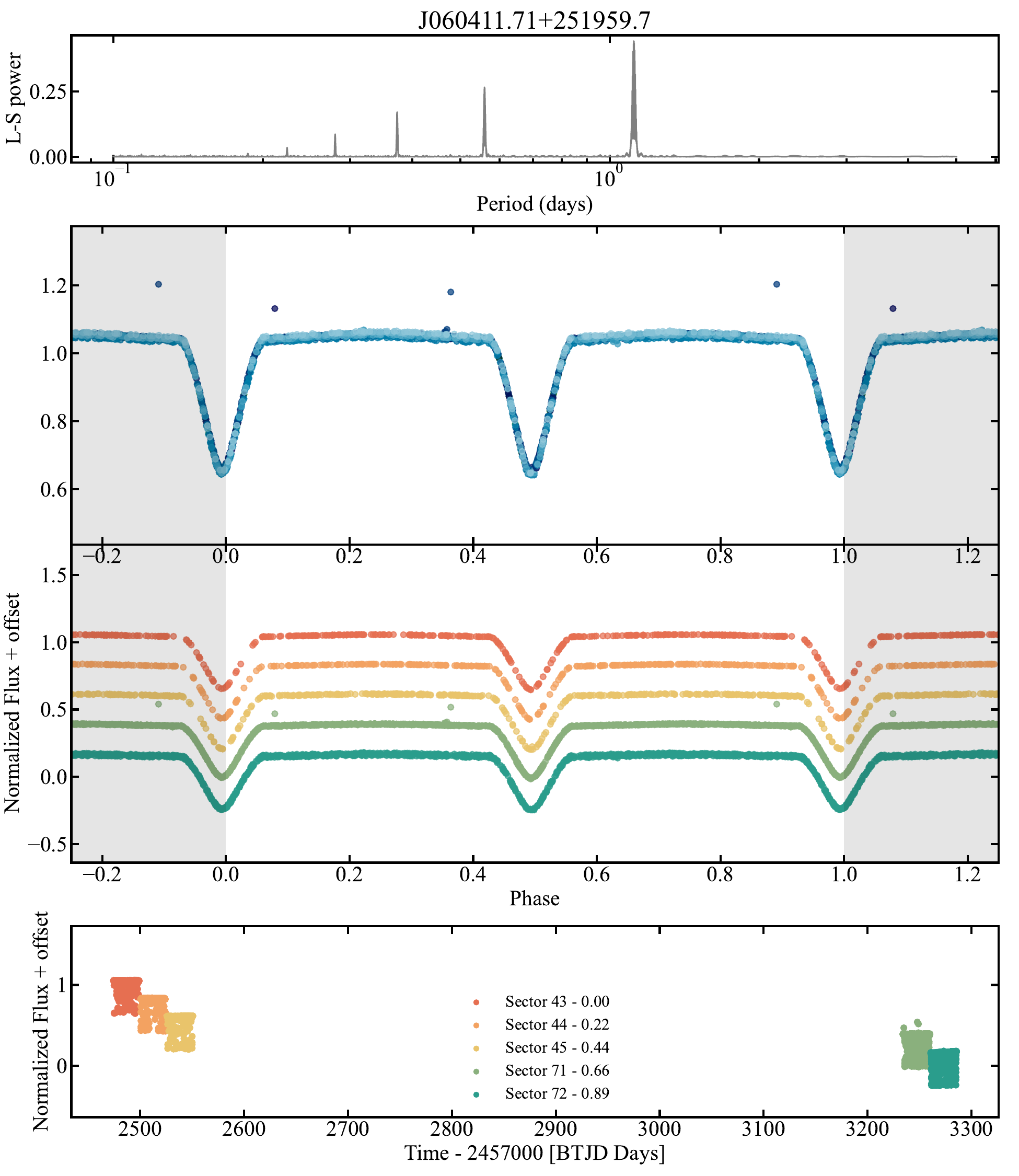}
\hspace{0.3in}
\includegraphics[scale=0.22]{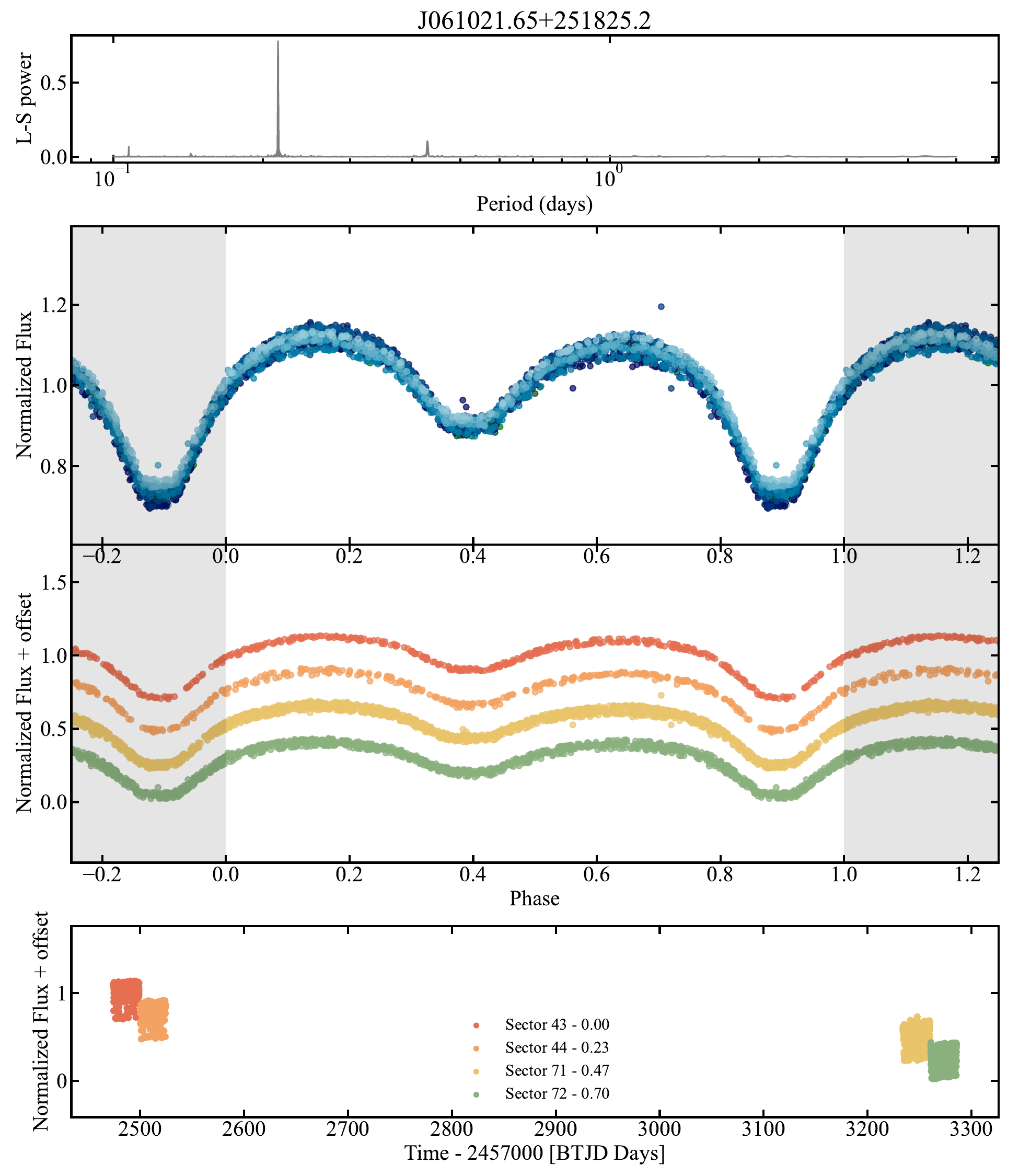}
\caption{Two examples of eclipsing type light curves (J060411.71+251959.7 on the left side and J061021.65+251825.2 on the right side). For each figure, the bottom-most row represents the observation from the entire TESS mission, with different mission phases separated by offsets. The third row displays the folded light curve using peak periods from the power spectrum. The second row presents the folded light curve without offset divisions and with double period copied. The top row depicts the power spectrum.
\label{Tess_fig}}
\end{figure}

\begin{figure}[h]
\centering
\includegraphics[scale=0.22]{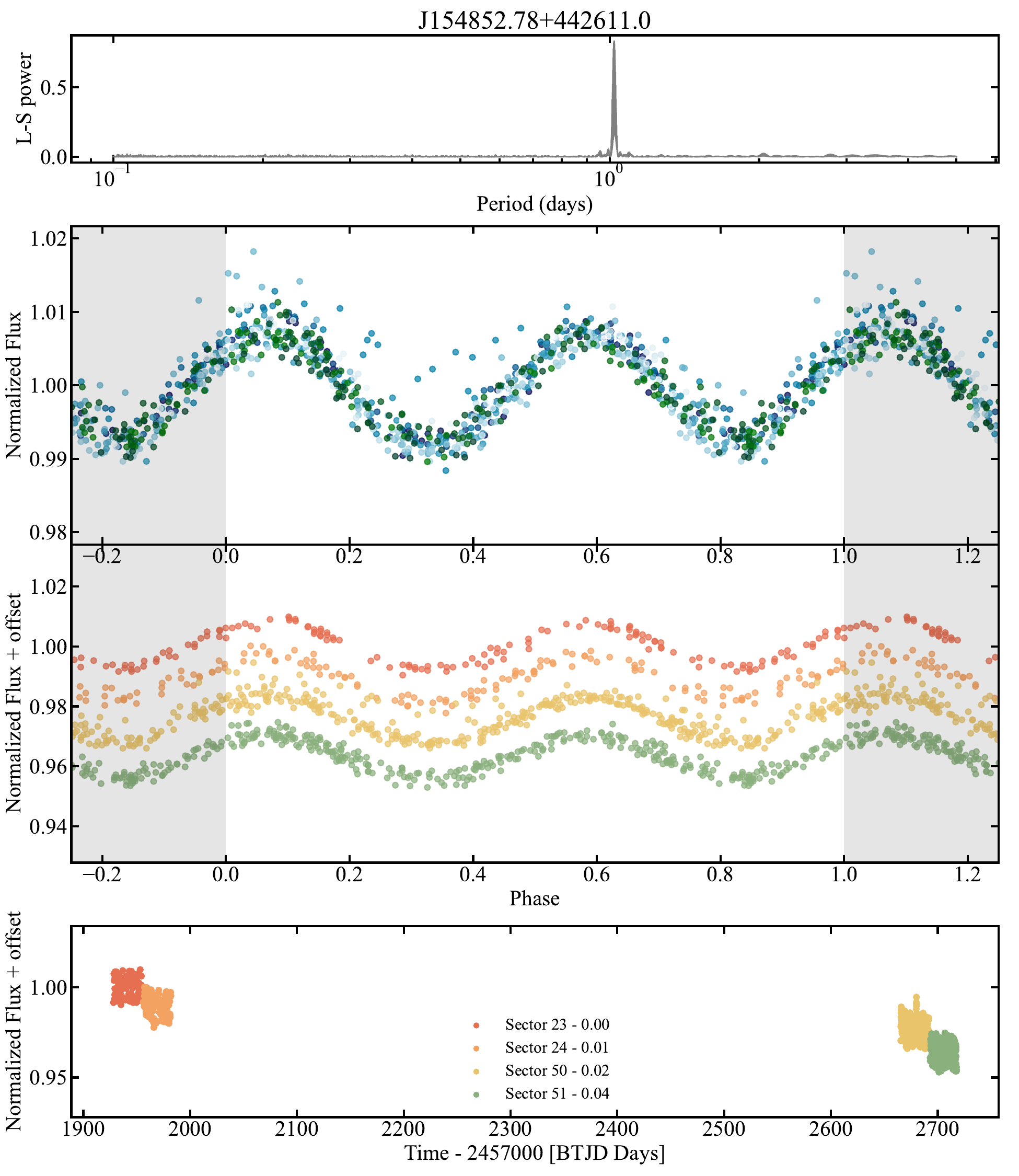}
\hspace{0.3in}
\includegraphics[scale=0.22]{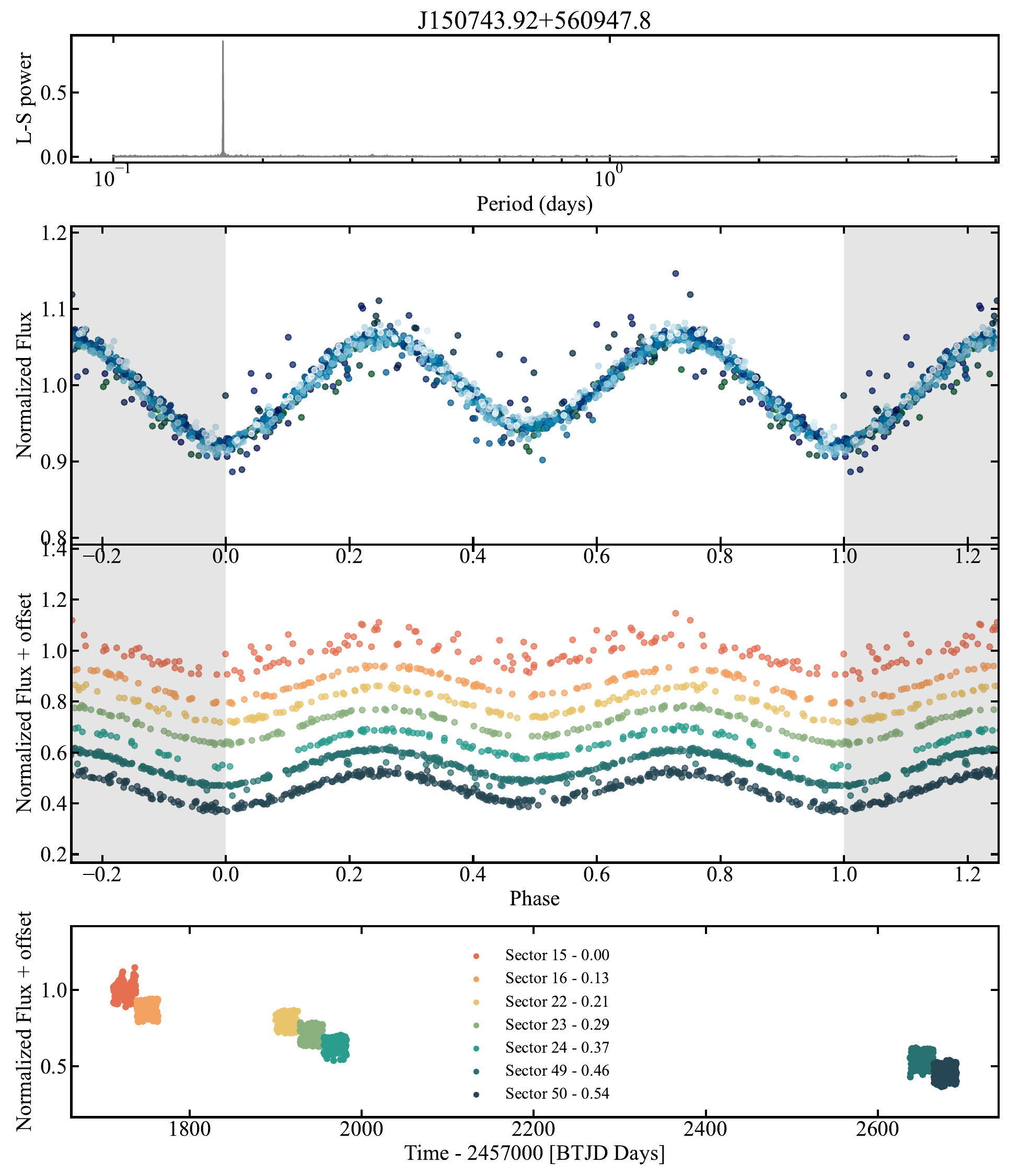}
\caption{Two examples of ellipsoidal light curves similar to Figure \ref{Tess_fig}, which are J154852.78+442611.0 (left) and J150743.92+560947.8 (right).
\label{ess_fig}}
\end{figure}

\clearpage

\section{LAMOST Spectroscopy}
\label{spec_CCF}
We use the cross-correlation function (CCF) to measure the relative radial velocities of the LAMOST spectra. The template is the spectrum with the highest signal-to-noise ratio for each target. To verify that the systems are single-lined spectroscopic binaries, we checked whether there are additional peaks in the CCF profile indicates the presence of extra components in the spectrum. We measure all available spectra for each target to compute the CCF. We check spectra which are chosen to be mostly near the quadrature phases. In these phases, if there are two components, they would be easily detected in the velocity space. As shown in the following figure, it is evident from the CCF that J061400 is likely to be a double-lined spectroscopic binary, while J075855 is a single-lined spectroscopic binary. 
 
\begin{figure}[h]
\centering
\includegraphics[scale=0.42]{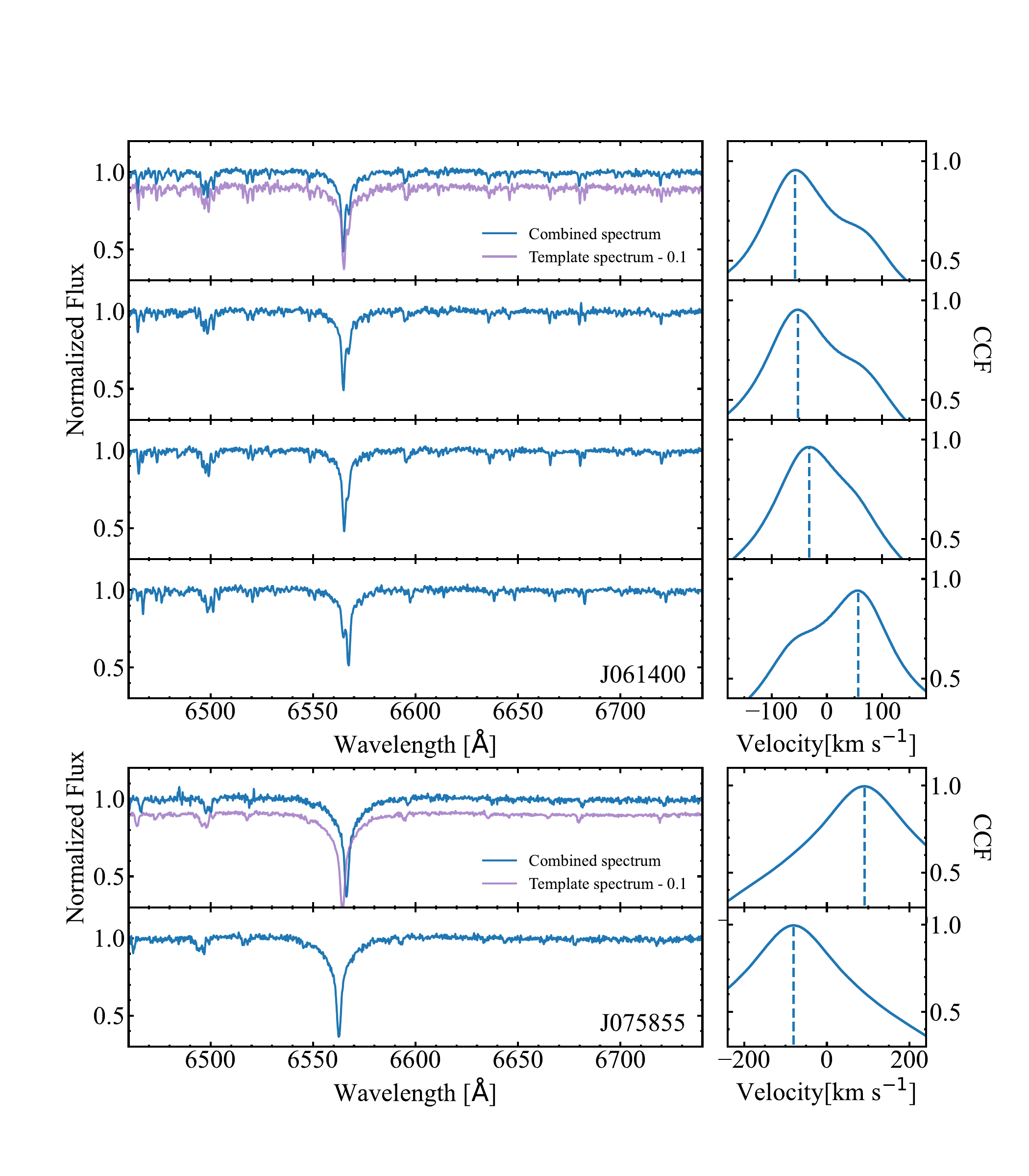}
\caption{LAMOST spectra (left panels) and CCF (right panels) for J061400 (upper panels) and J075855 (lower panels). The uppermost panel in J061400 shows the combined spectrum (blue) and a template spectrum (red) used to measure the CCF profile. The dashed lines mark the peak in the CCF panels. 
\label{spec}}
\end{figure}

\clearpage
\section{The mass distributions}
\label{mass_all}
  \begin{figure}[h]
  \label{full_mass}
\centering
\includegraphics[scale=0.6]{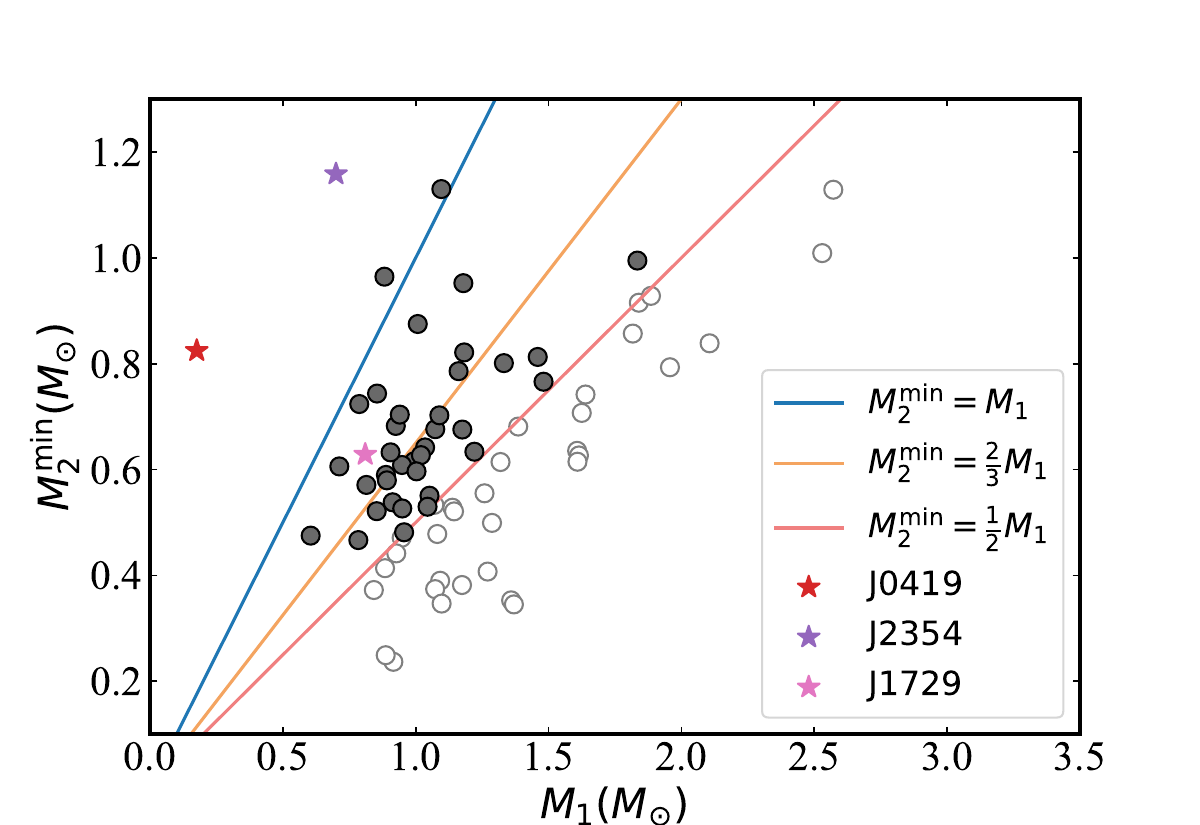}
\caption{The minimum mass of the unseen companion at the maximum inclination ($i = 90^{\circ}$) similar to Figure \ref{mass}. Three solid lines represent the cases $M_2^\mathrm{min} = M_1$, $M_2^\mathrm{min} = 2 M_1/3$ and $M_2^\mathrm{min} = M_1/2$, respectively. The grey circles in the graph indicate 36 targets with lower limit of mass ratio ($M_2^\mathrm{min} / M_1$) exceeding 1/2, while the white circles are others.
\label{mass2}}
\end{figure}

  \section{Full Selected Samples}
  \label{full_tab}

  \startlongtable
  \begin{deluxetable*}{ccccccccccc}
    \tablecaption{Characteristics of 89 selected candidates
    \label{suptable}}
    \tablewidth{0.99\textwidth}
    \tabletypesize{\scriptsize}
    \tablehead{
      \colhead{Number} & \colhead{Designation} & \colhead{Source ID}      & \colhead{\neff}        & \colhead{$\Delta \rvi$} & \colhead{\Porb} & \colhead{LC} & \colhead{$K_1$} & \colhead{\massfunc} & \colhead{$M_1$}        & \colhead{$M_2^\mathrm{min}$}\\
      \colhead{}          & \colhead{}            & \colhead{}    & \colhead{(1)}       & \colhead{(2)}       & \colhead{(3)}           & \colhead{(4)}     & \colhead{(5)} & \colhead{(6)}  & \colhead{(7)}& \colhead{(8)}\\
      \colhead{}          & \colhead{}            & \colhead{}    & \colhead{(days)} & \colhead{(km s$^{-1}$)} & \colhead{(days)} & \colhead{} & \colhead{(km s$^{-1}$)}     & \colhead{(\msun)}   & \colhead{(\msun)}   & \colhead{(\msun)}             
    }
    \startdata
        1     & J041920.07+072545.4 & 410015 & 31    & 418.7 & 0.61  &   ELL & 208.6 & 0.56  &       &  \\
    2     & J235456.76+335625.7 & 1129096 & 5     & 426.5 & 0.48  &   ELL & 210.5 & 0.455 &       &  \\
    3     & J134417.15+520307.2 & 818528 & 5     & 130.55 & 3.21  &       & 95.04 & 0.291 & 1.10  & 1.13  \\
    4     & J102740.93+384018.9 & 697652 & 11    & 126.44 & 2.66  &   ELL & 67.87 & 0.263 & 0.88  & 0.96  \\
    5     & J122924.56+473128.5 & 771613 & 7     & 185.76 & 2.23  &   ELL & 93.9  & 0.19  & 1.18  & 0.95  \\
    6     & J060826.18+211334.1 & 400747 & 6     & 141.74 & 3.87  &       & 78.42 & 0.189 & 1.01  & 0.87  \\
    7     & J122302.07+482910.8 & 767594 & 10    & 189.98 & 1.57  &   ELL & 102.96 & 0.166 & 0.79  & 0.72  \\
    8     & J060030.98+290855.0 & 387387 & 11    & 447.04 & 0.15  &   ELL & 210.61 & 0.162 &       &  \\
    9     & J085304.59+132032.3 & 616838 & 17    & 249.3 & 0.79  &   ELL & 128.69 & 0.161 & 0.85  & 0.74  \\
    10    & J060649.24+213233.3 & 397914 & 12    & 157.38 & 2.77  &   ELL & 78.87 & 0.14  &       &  \\
    11    & J060907.13+222550.6 & 401948 & 11    & 128.91 & 2.52  &       & 82.49 & 0.138 & 1.18  & 0.82  \\
    12    & J092323.52+421819.5 & 651359 & 15    & 134.79 & 3.64  &   ELL & 68.54 & 0.129 & 0.94  & 0.70  \\
    13    & J090332.95+425639.8 & 630705 & 35    & 210.12 & 0.89  &   ELL & 59.07 & 0.128 & 1.16  & 0.79  \\
    14    & J230854.08+355132.4 & 1095399 & 9     & 207.93 & 0.95  &   ELL & 110.92 & 0.128 & 0.71  & 0.61  \\
    15    & J044826.19+223906.2 & 296802 & 7     & 154.56 & 2.47  &   ELL & 78.59 & 0.124 &       &  \\
    16    & J063350.35+220056.0 & 447835 & 7     & 235.13 & 0.69  &       & 123.88 & 0.123 & 1.83  & 0.99  \\
    17    & J155751.89+435021.4 & 884025 & 6     & 175.58 & 1.87  &       & 86.54 & 0.123 & 0.93  & 0.68  \\
    18    & J172900.17+652952.8 & 905386 & 1     & 94.8  & 0.6   &   ELL & 93.87 & 0.123 &       &  \\
    19    & J075855.51+384540.7 & 550320 & 15    & 177.7 & 1.35  &   ELL & 92.97 & 0.113 & 1.33  & 0.80  \\
    20    & J103613.90+071115.6 & 703271 & 10    & 113.53 & 1.5   &       & 61.41 & 0.108 & 1.09  & 0.70  \\
    21    & J061921.71+224432.3 & 422288 & 10    & 163.02 & 2.56  &   ELL & 67.76 & 0.107 & 0.91  & 0.63  \\
    22    & J061125.78+115043.5 & 406053 & 6     & 114.71 & 1.93  &       & 82.93 & 0.105 & 2.57  & 1.13  \\
    23    & J011726.82+452013.2 & 64557 & 7     & 163.61 & 2.08  &       & 78.89 & 0.104 & 1.46  & 0.81  \\
    24    & J022924.20+581721.3 & 135083 & 6     & 109.53 & 5.07  &   ELL & 53.93 & 0.101 & 1.89  & 0.93  \\
    25    & J202309.48+401916.5 & 1010372 & 7     & 134.48 & 1.18  &   ELL & 94.04 & 0.101 & 1.84  & 0.92  \\
    26    & J060537.69+222546.9 & 395839 & 13    & 136.43 & 3.25  &   ELL & 67.55 & 0.101 & 1.07  & 0.68  \\
    27    & J085553.20+181128.2 & 621071 & 15    & 133.92 & 3.26  &   ELL & 66.68 & 0.097 & 0.81  & 0.57  \\
    28    & J033013.01+591627.5 & 193088 & 7     & 172.03 & 0.77  &   ELL & 120.59 & 0.094 & 1.04  & 0.64  \\
    29    & J035557.04+235457.0 & 223483 & 13    & 135.03 & 2.72  &   ELL & 69.98 & 0.094 & 0.89  & 0.59  \\
    30    & J155244.61+461437.5 & 881769 & 13    & 117.01 & 4.11  &   ELL & 60.9  & 0.093 & 0.95  & 0.61  \\
    31    & J154452.52+505346.9 & 878260 & 10    & 236.82 & 0.46  &   ELL & 120.03 & 0.092 & 0.60  & 0.48  \\
    32    & J064126.68+234312.8 & 463302 & 18    & 182.78 & 1.24  &       & 87.94 & 0.091 &       &  \\
    33    & J050047.77+493317.9 & 314676 & 9     & 120.79 & 3.93  &   ELL & 60.4  & 0.091 & 1.02  & 0.63  \\
    34    & J084919.55+181050.1 & 611176 & 6     & 126.28 & 3.34  &   ELL & 67.4  & 0.09  & 1.18  & 0.68  \\
    35    & J104756.16+402735.3 & 712440 & 15    & 153.03 & 2.08  &   ELL & 76.23 & 0.09  & 0.89  & 0.58  \\
    36    & J091434.79+425744.2 & 642865 & 20    & 117.69 & 4.28  &       & 61.71 & 0.09  & 0.99  & 0.62  \\
    37    & J034813.39+250555.6 & 214231 & 17    & 130   & 3.93  &   ELL & 61.74 & 0.089 & 1.48  & 0.77  \\
    38    & J064717.33+243633.7 & 473755 & 16    & 128.46 & 2.91  &   ELL & 68.91 & 0.088 & 1.82  & 0.86  \\
    39    & J112713.81+003407.1 & 737276 & 7     & 109.38 & 4.44  &   ELL & 56.75 & 0.083 & 1.00  & 0.60  \\
    40    & J044422.50+483630.6 & 291282 & 17    & 123.93 & 3.22  &   ELL & 61.74 & 0.082 &       &  \\
    41    & J063552.55+184233.7 & 451888 & 6     & 149.44 & 1.96  &   ELL & 74.73 & 0.082 & 2.53  & 1.01  \\
    42    & J032320.58+521217.3 & 185497 & 14    & 133.14 & 2.74  &   ELL & 69.05 & 0.078 & 4.27  & 1.35  \\
    43    & J104734.61+075525.0 & 712171 & 16    & 116.39 & 4.1   &       & 56.02 & 0.075 & 0.85  & 0.52  \\
    44    & J145742.06+385246.4 & 852175 & 16    & 110.86 & 4.22  &   ELL & 54.2  & 0.074 & 1.22  & 0.63  \\
    45    & J112022.64+030629.2 & 732821 & 6     & 117.04 & 3.05  &   ELL & 65.16 & 0.074 & 1.39  & 0.68  \\
    46    & J105637.88+104446.8 & 718646 & 9     & 162.64 & 1.37  &   ELL & 80.28 & 0.074 & 0.91  & 0.54  \\
    47    & J050627.51+443221.7 & 320573 & 6     & 124.89 & 2.91  &   ELL & 62.93 & 0.072 & 1.64  & 0.74  \\
    48    & J230328.86+365245.5 & 1092454 & 7     & 108.04 & 3.96  &   ELL & 56.62 & 0.07  &       &  \\
    49    & J061213.85+213955.7 & 407534 & 16    & 114.8 & 4.59  &       & 52.38 & 0.068 & 2.11  & 0.84  \\
    50    & J001323.45+574834.3 & 15183 & 9     & 118.05 & 2.82  &       & 60.96 & 0.068 &       &  \\
    51    & J011110.49+092102.5 & 57176 & 13    & 107.71 & 4.26  &   ELL & 54.44 & 0.067 & 0.95  & 0.53  \\
    52    & J035912.12+570228.8 & 227541 & 9     & 137.63 & 1.38  &   ELL & 77.48 & 0.066 & 1.96  & 0.79  \\
    53    & J010421.68+041336.5 & 48956 & 8     & 196.25 & 0.57  &       & 104.86 & 0.065 &       &  \\
    54    & J064044.85+235615.5 & 461798 & 24    & 117.5 & 3.15  &   ELL & 58.94 & 0.065 & 1.62  & 0.71  \\
    55    & J142953.75+460353.8 & 841276 & 9     & 232.55 & 0.59  &       & 95.67 & 0.065 & 0.78  & 0.47  \\
    56    & J080602.82+410506.1 & 558589 & 10    & 132.27 & 1.72  &       & 60.96 & 0.065 & 1.05  & 0.55  \\
    57    & J023639.27+553554.8 & 143047 & 6     & 138.94 & 0.89  &   ELL & 86.7  & 0.062 &       &  \\
    58    & J030357.53+545157.6 & 167876 & 17    & 118.98 & 2.64  &       & 63.53 & 0.062 & 1.32  & 0.61  \\
    59    & J011256.97+451943.3 & 59447 & 7     & 115.02 & 2.41  &   ELL & 63.14 & 0.062 &       &  \\
    60    & J034121.29+573315.0 & 205866 & 6     & 125.71 & 1.1   &   ELL & 65.06 & 0.06  & 1.04  & 0.53  \\
    61    & J010641.31+025621.7 & 51463 & 9     & 177.95 & 0.87  &   ELL & 84.44 & 0.059 & 1.07  & 0.53  \\
    62    & J193028.19+440007.2 & 1004002 & 15    & 118.77 & 2.57  &   ELL & 60.96 & 0.057 &       &  \\
    63    & J225115.59+345157.4 & 1085828 & 9     & 138.71 & 1.74  &   ELL & 67.72 & 0.054 & 0.96  & 0.48  \\
    64    & J231017.65+334202.6 & 1096283 & 15    & 102.14 & 3.61  &       & 63.64 & 0.053 & 1.14  & 0.53  \\
    65    & J125038.89+535228.9 & 783510 & 7     & 100.42 & 3.94  &   ELL & 51.24 & 0.052 & 0.95  & 0.47  \\
    66    & J064742.91+244600.4 & 474426 & 11    & 137.94 & 1.54  &   ELL & 69.31 & 0.052 & 1.26  & 0.56  \\
    67    & J060429.52+344052.7 & 393673 & 7     & 101.57 & 2.22  &       & 64.99 & 0.051 & 1.61  & 0.64  \\
    68    & J092306.86+431939.7 & 651103 & 34    & 165.83 & 1.31  &   ELL & 72.04 & 0.051 &       &  \\
    69    & J085125.29+120256.4 & 614307 & 16    & 123.86 & 2.82  &   ELL & 61.4  & 0.051 & 1.14  & 0.52  \\
    70    & J225434.80+344144.2 & 1087587 & 12    & 156.5 & 1.13  &   ELL & 75.89 & 0.049 & 1.61  & 0.63  \\
    71    & J064109.34+233700.5 & 462696 & 20    & 144.73 & 1.48  &   ELL & 69.19 & 0.047 & 1.61  & 0.61  \\
    72    & J203510.62+422022.8 & 1013286 & 7     & 113.57 & 2.45  &       & 57.17 & 0.046 &       &  \\
    73    & J034544.41+241313.1 & 211191 & 14    & 101.42 & 3.31  &   ELL & 51.63 & 0.046 & 0.93  & 0.44  \\
    74    & J043432.74+272850.3 & 280833 & 8     & 113.24 & 2.31  &   ELL & 61.27 & 0.045 & 1.08  & 0.48  \\
    75    & J074928.34+392312.9 & 539298 & 20    & 101.39 & 3.04  &   ELL & 51.53 & 0.042 & 0.88  & 0.41  \\
    76    & J085056.28+131737.6 & 613574 & 5     & 115.44 & 0.68  &   ELL & 81.8  & 0.039 & 1.29  & 0.50  \\
    77    & J064601.31+235035.7 & 471648 & 19    & 127.44 & 1.25  &       & 64.52 & 0.035 & 0.84  & 0.37  \\
    78    & J203932.62+405357.6 & 1014996 & 7     & 180.12 & 2.26  &   ELL & 52.36 & 0.033 &       &  \\
    79    & J041640.72+485628.8 & 252081 & 11    & 106.75 & 1.97  &   ELL & 53.33 & 0.03  &       &  \\
    80    & J084000.87+183844.4 & 599891 & 9     & 105.69 & 1.29  &   ELL & 59.68 & 0.027 & 1.09  & 0.39  \\
    81    & J080832.08+423555.1 & 561935 & 19    & 121.27 & 1.17  &   ELL & 60.88 & 0.025 & 1.07  & 0.37  \\
    82    & J222623.64+302804.0 & 1074150 & 13    & 120.21 & 2.1   &   ELL & 47.67 & 0.024 & 1.27  & 0.41  \\
    83    & J031618.78+324339.4 & 178887 & 8     & 107.94 & 1.54  &       & 52.91 & 0.023 & 1.17  & 0.38  \\
    84    & J083756.41+123453.1 & 597848 & 20    & 116.57 & 1.44  &   ELL & 52.9  & 0.02  & 1.10  & 0.35  \\
    85    & J222256.75+305648.8 & 1071843 & 13    & 107.11 & 2.37  &   ELL & 39.97 & 0.015 & 1.36  & 0.35  \\
    86    & J064220.56+240502.9 & 465138 & 19    & 106.66 & 2     &   ELL & 41.57 & 0.014 & 1.37  & 0.35  \\
    87    & J090436.48+434449.0 & 631750 & 16    & 96.78 & 1.22  &       & 46.16 & 0.012 & 0.89  & 0.25  \\
    88    & J103218.28+052318.7 & 700588 & 7     & 127.07 & 2.98  &       & 34.37 & 0.01  & 0.92  & 0.24  \\
    89    & J055058.33+285141.1 & 376806 & 7     & 106.92 & 2.22  &       & 49.79 & 0.005 &       &  \\
    \enddata
    \tablecomments{Column(1) The effective observation days in LAMOST DR10-MRS. Column(2) The maximum difference for effective mean radial velocity \rvi in LAMOST DR10-MRS. Column(3) The orbital period obtained from the radial velocity curve fitting. Column(4) The types of light curve, Ell stands for ellipsoidal modulation and blank denote no observably periodic change. Column(5) The semi-amplitude of the radial velocity obtained from the radial velocity curve fitting. Column(6)The mass function of the unseen companion calculated by Column(3) and Column(5). Column(7) The mass of visible star measured by \texttt{isochrones}. Column(8) The low mass limit of unseen companion estimated with inclination of $ i = 90^{\circ}$ 
\label{tab_lmf}}
  \end{deluxetable*}

\section{Massive candidates}
\label{rv_mass}
\begin{figure}[h]
\centering
\includegraphics[scale=0.25]{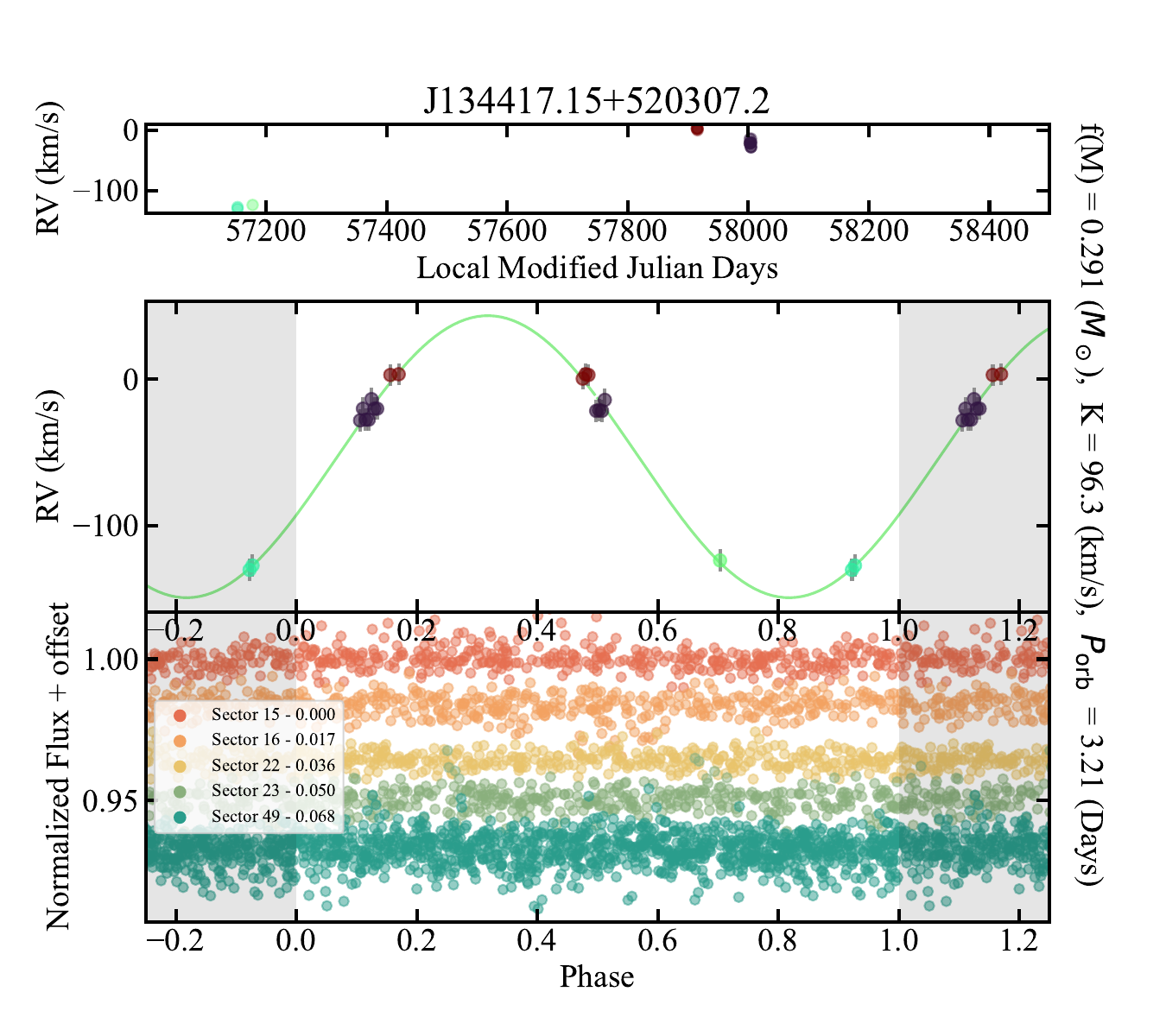}
\includegraphics[scale=0.25]{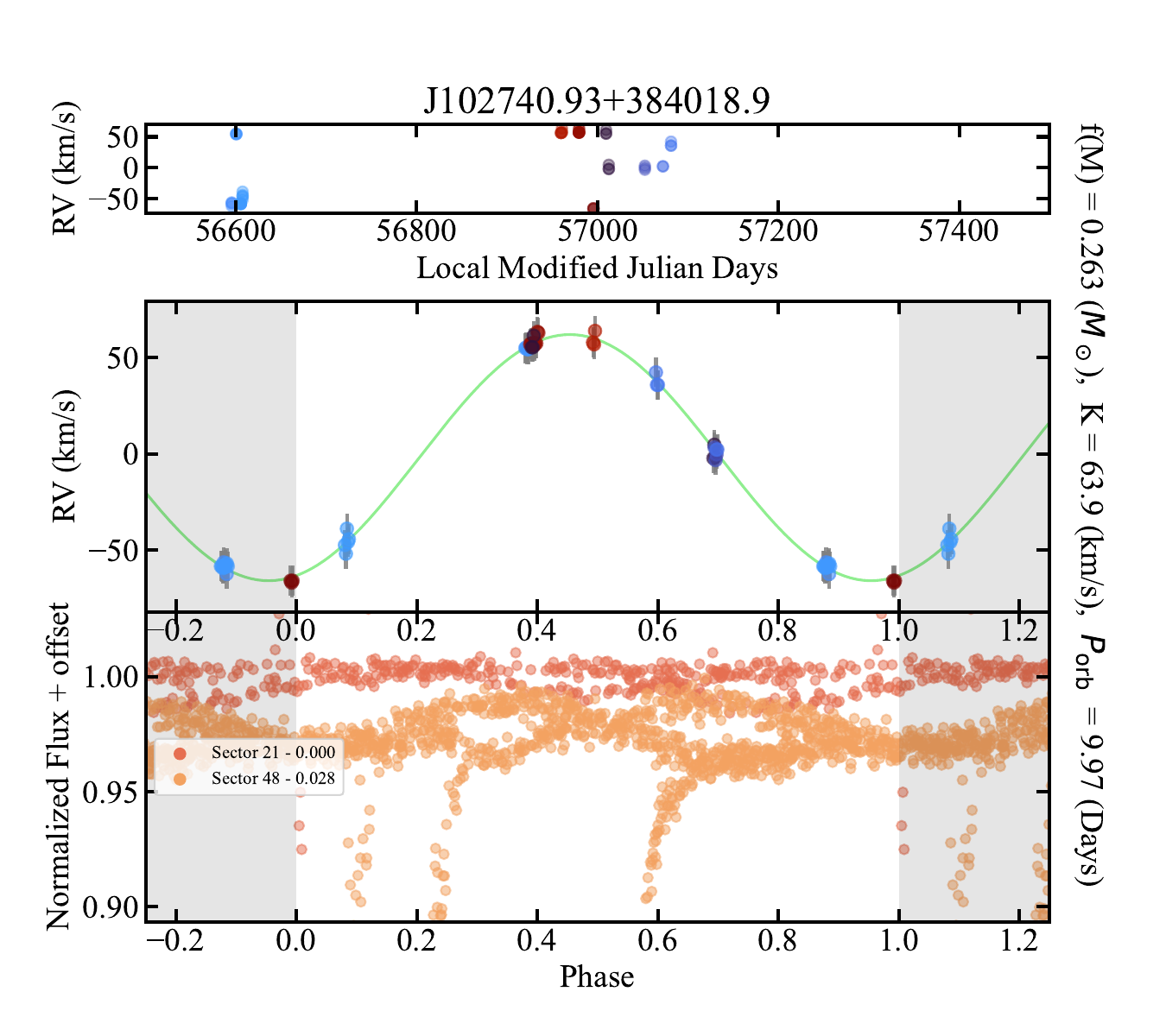}
\includegraphics[scale=0.25]{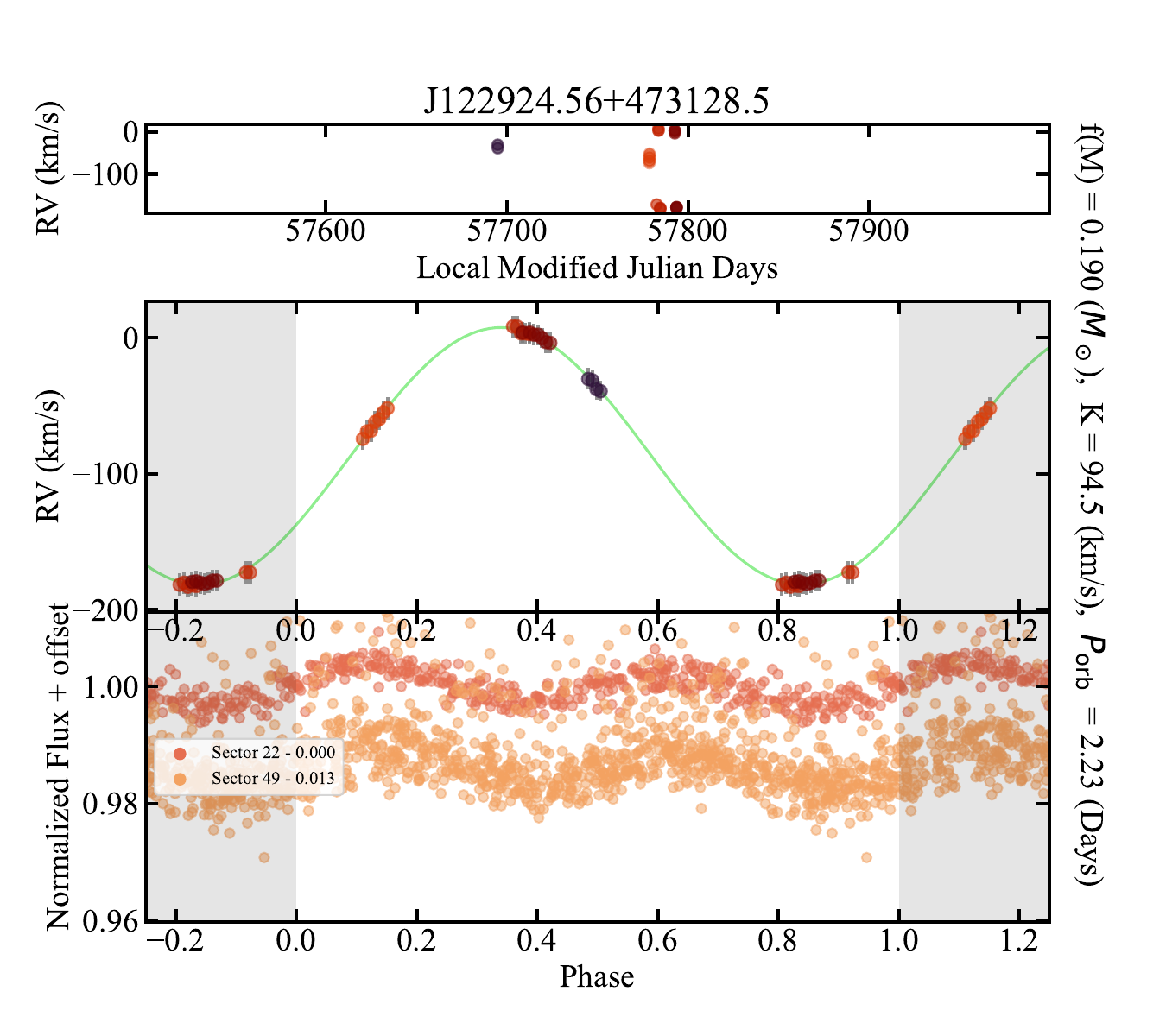}
\includegraphics[scale=0.25]{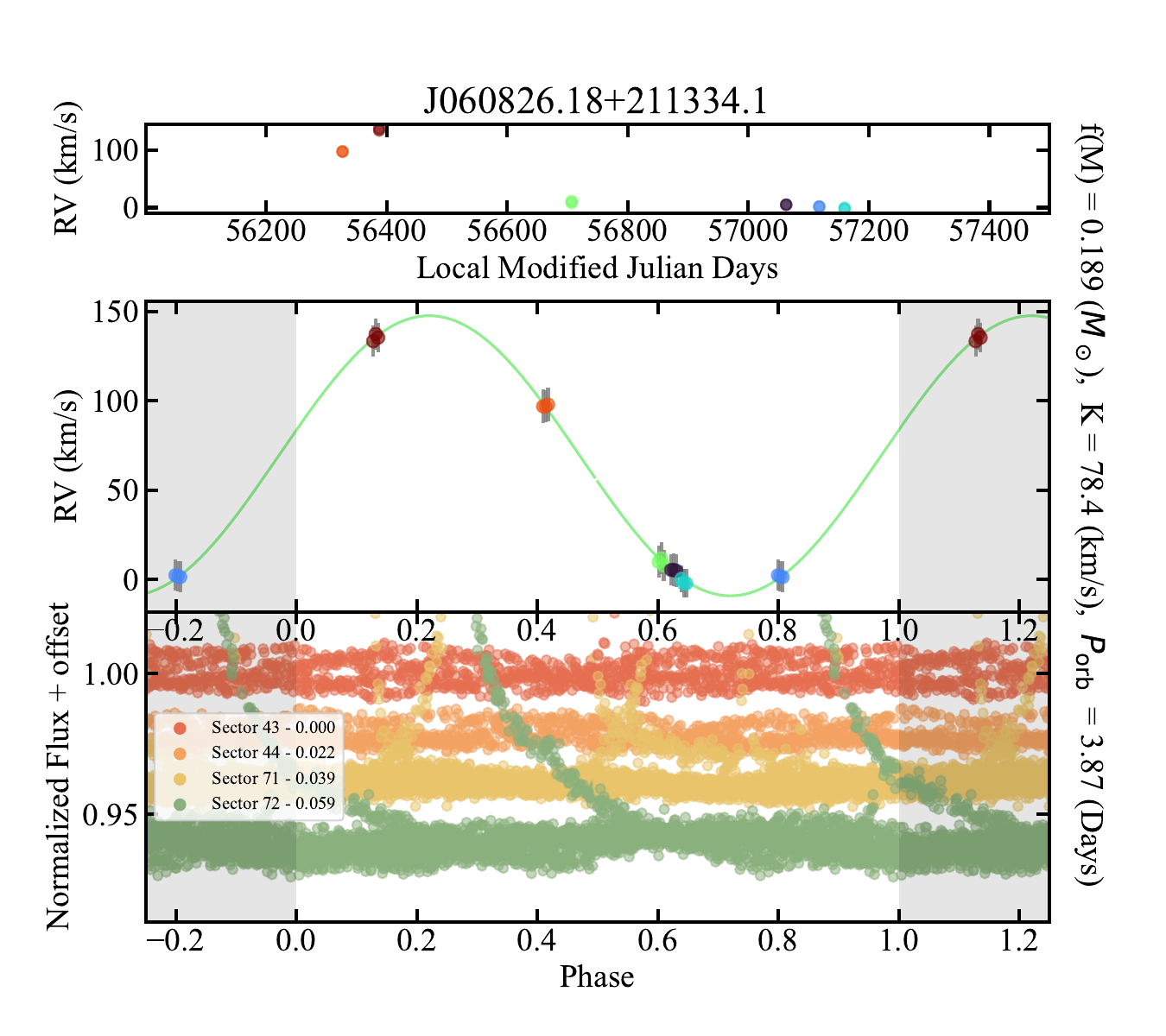}
\includegraphics[scale=0.25]{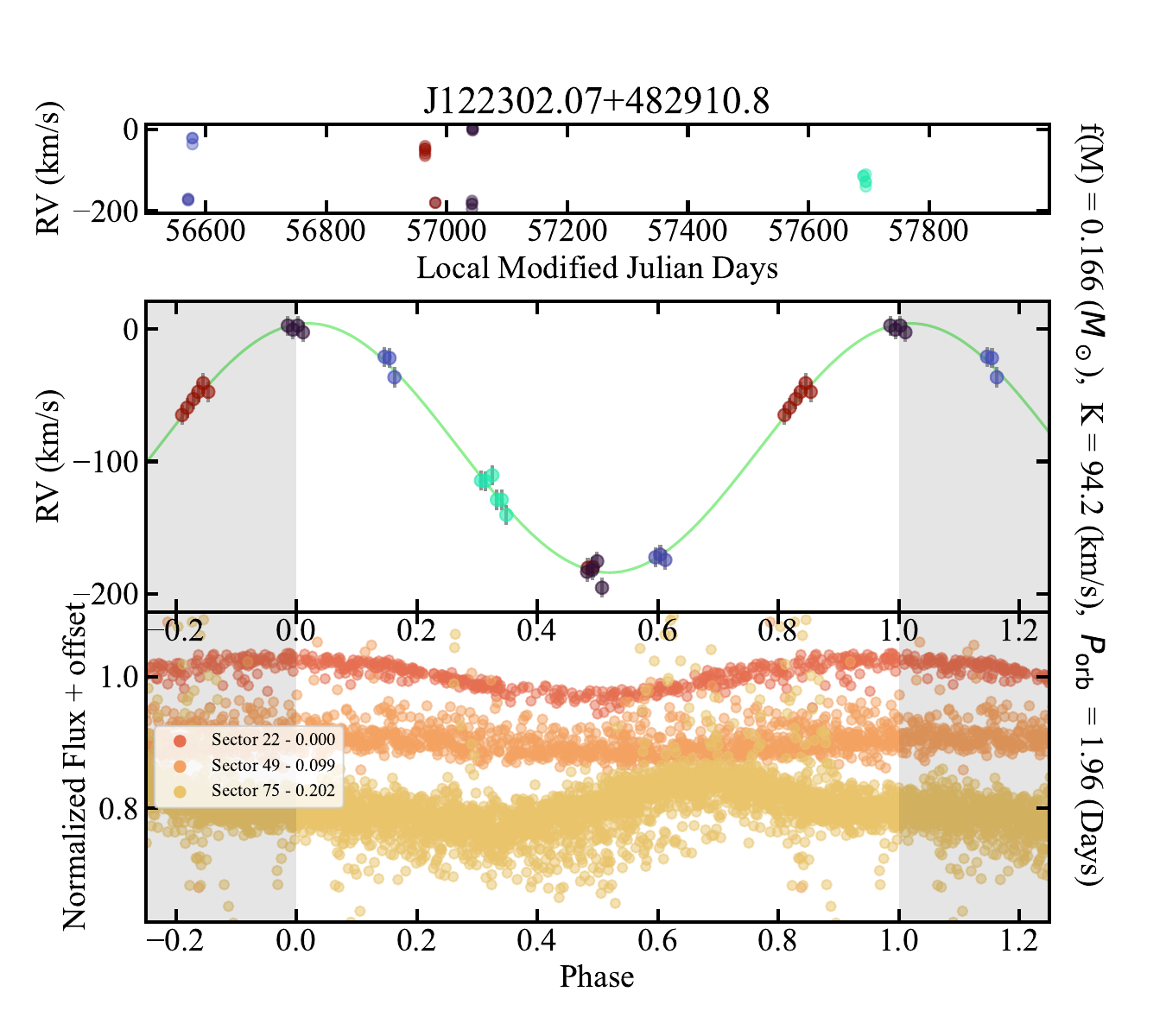}
\includegraphics[scale=0.25]{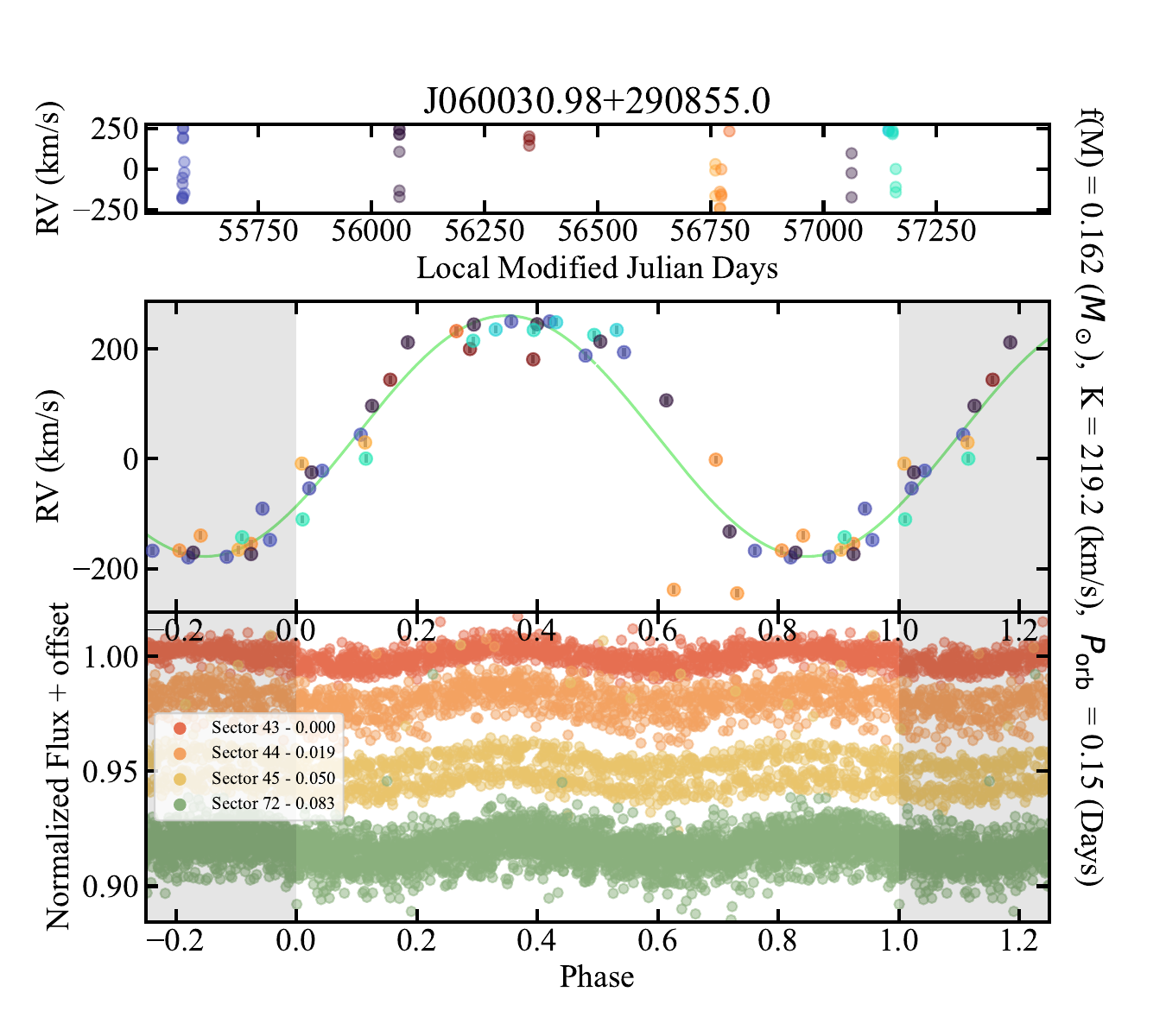}
\includegraphics[scale=0.25]{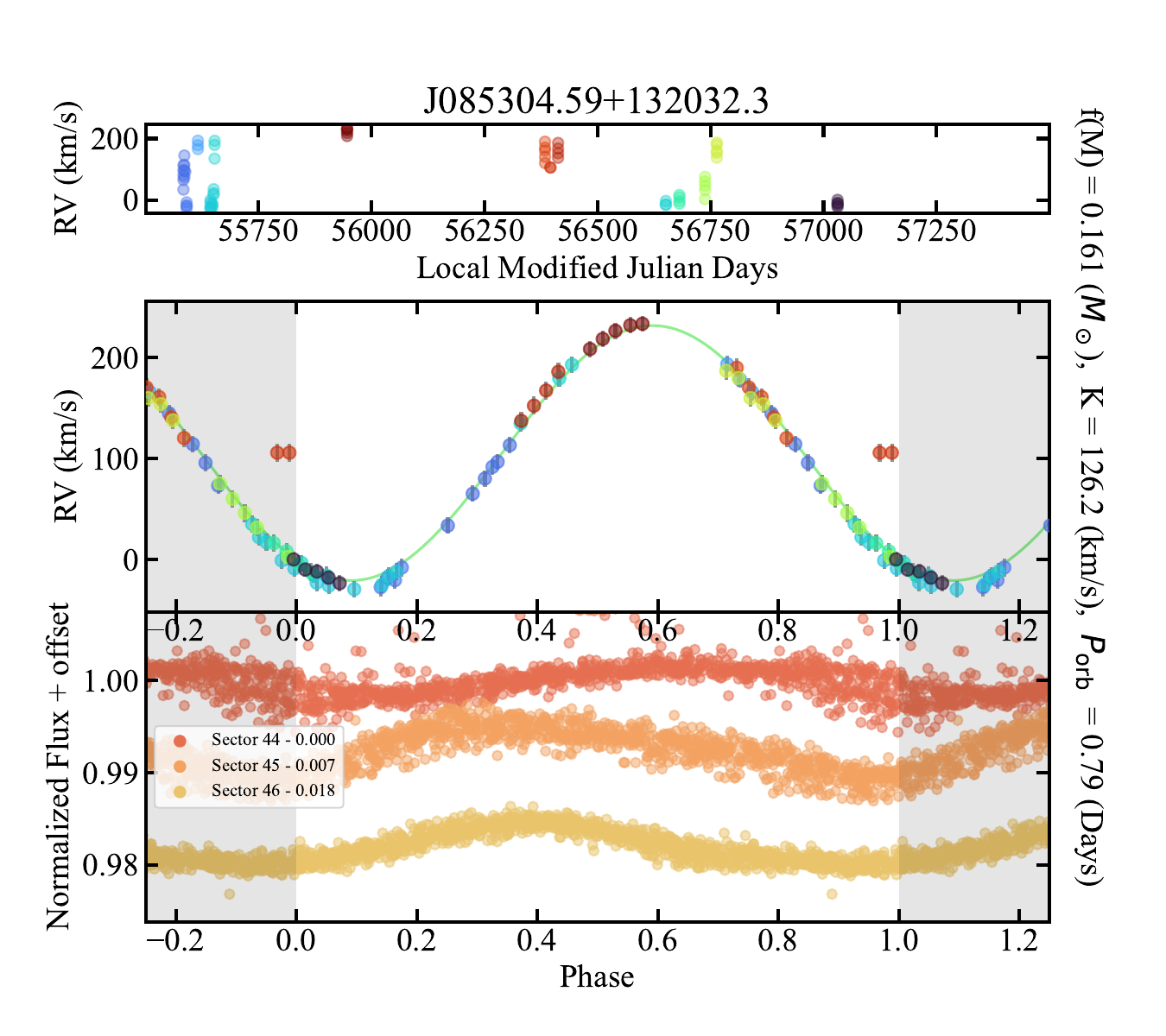}
\includegraphics[scale=0.25]{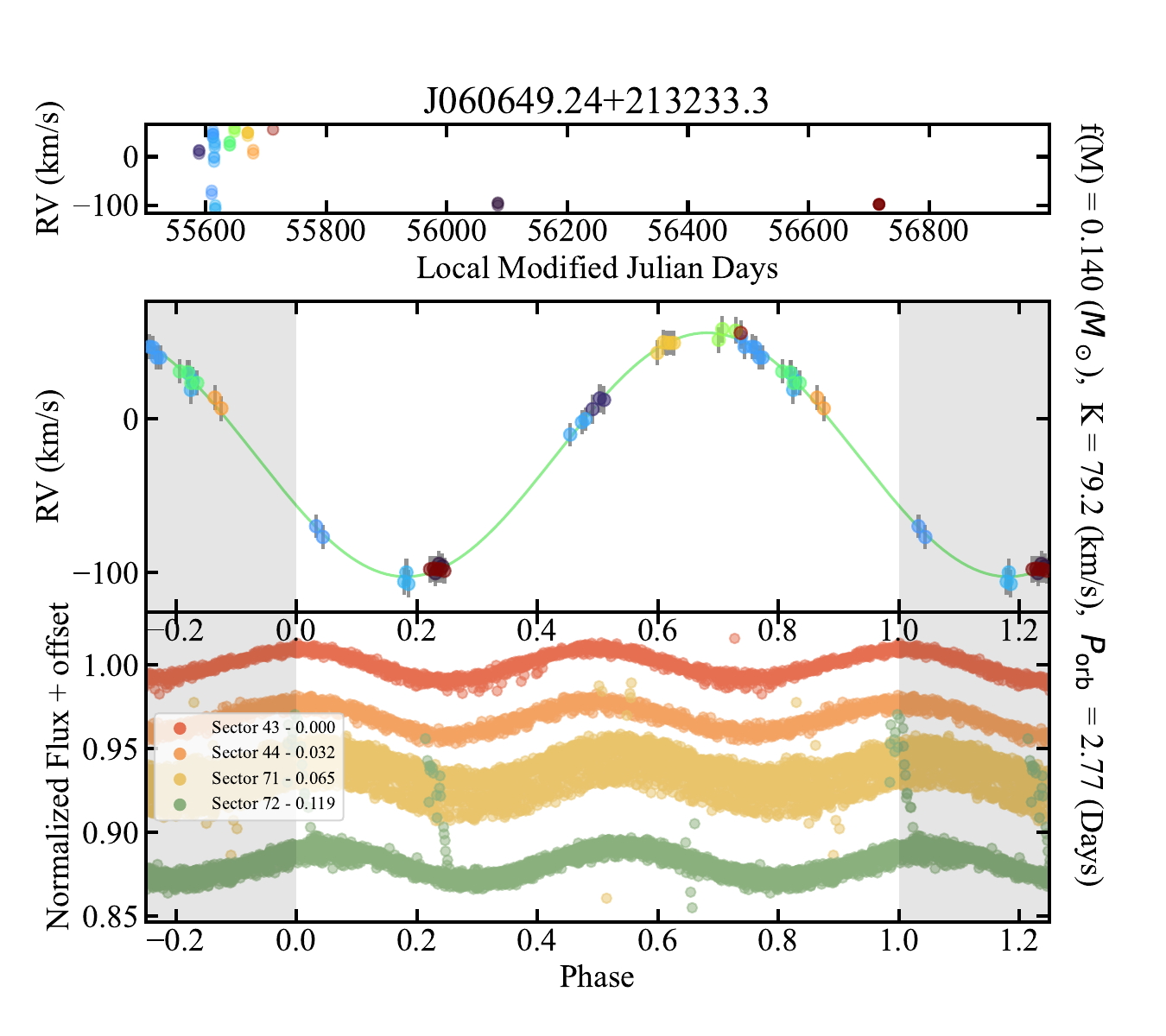}
\includegraphics[scale=0.25]{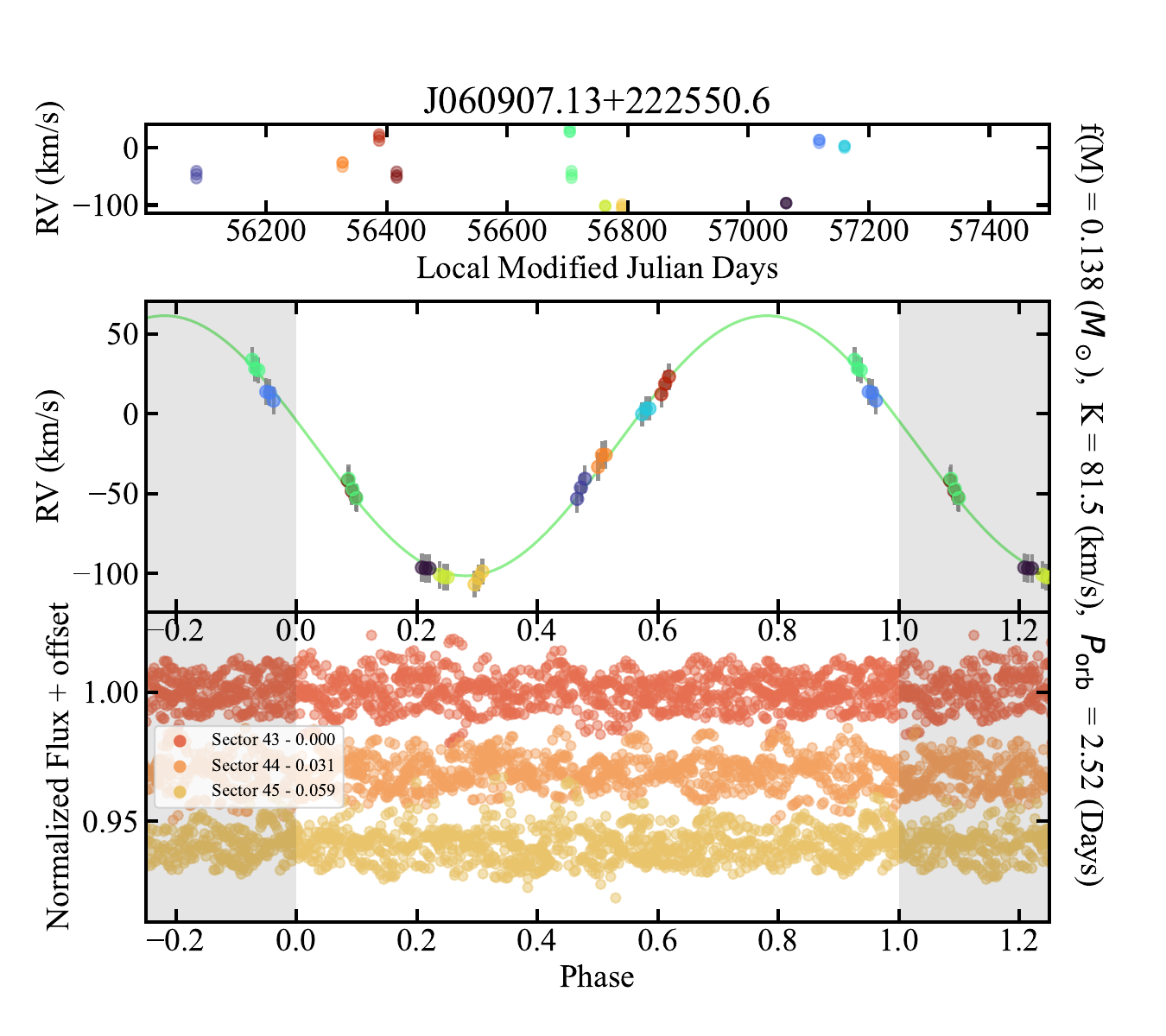}
\includegraphics[scale=0.25]{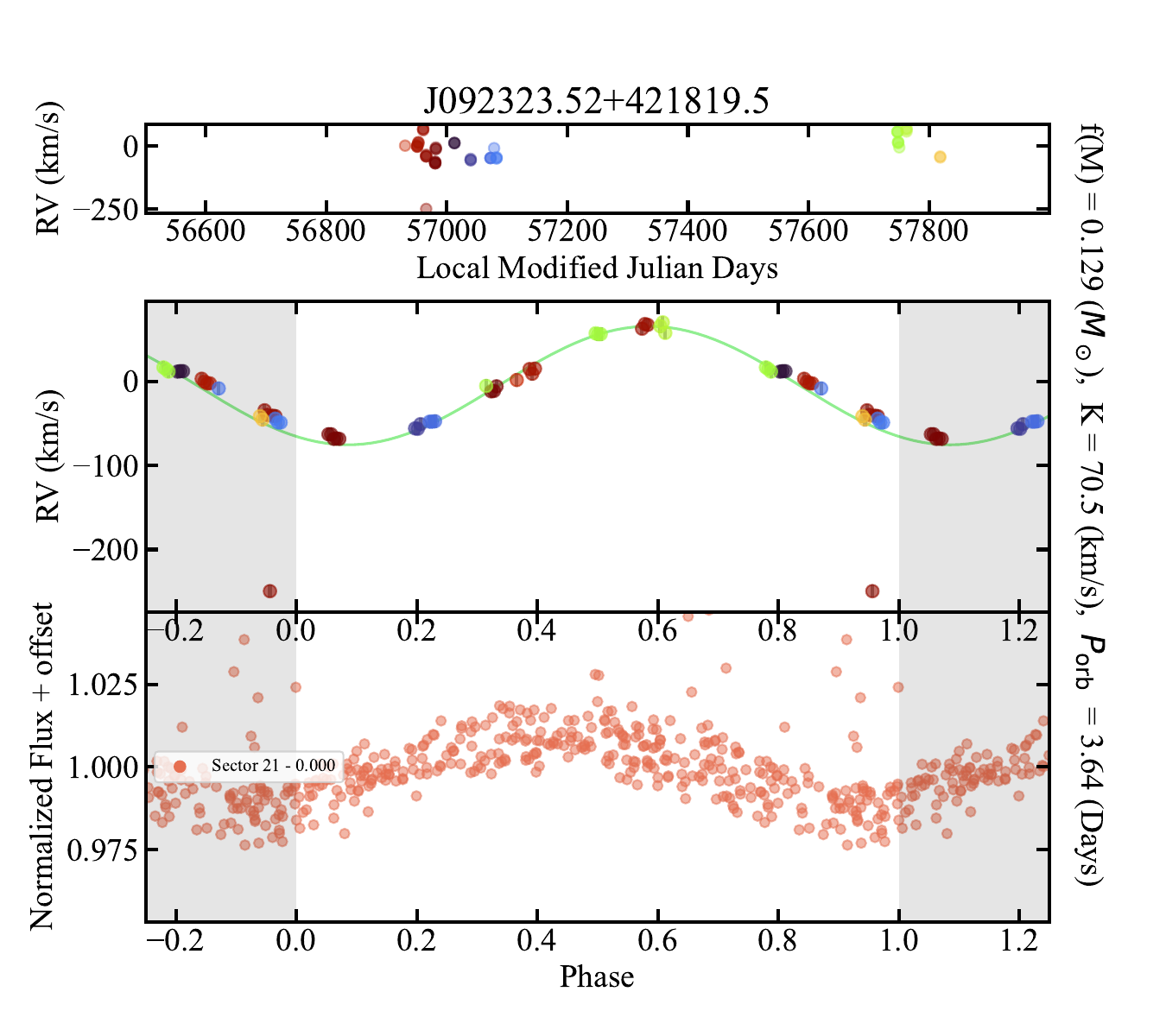}
\includegraphics[scale=0.25]{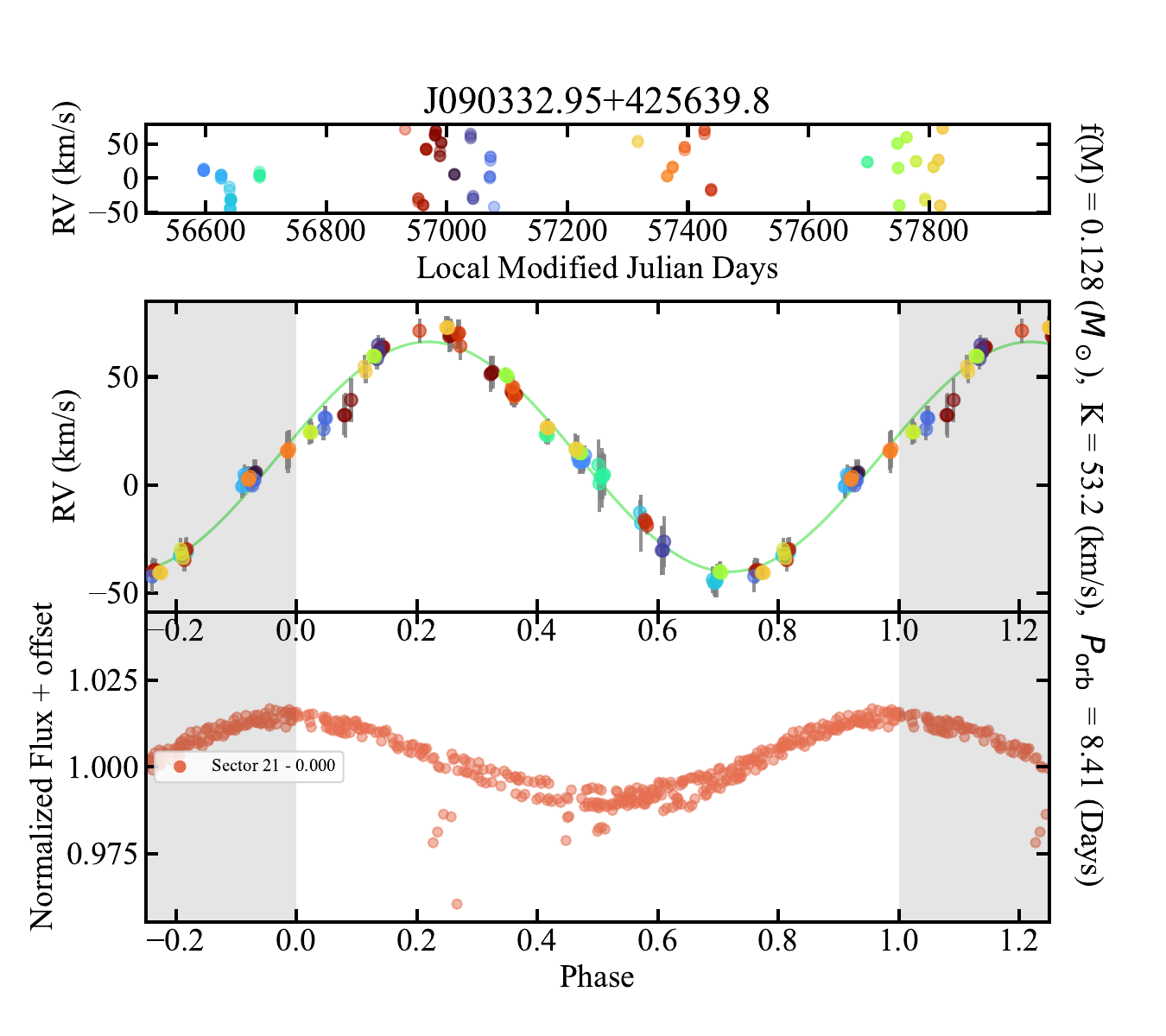}
\includegraphics[scale=0.25]{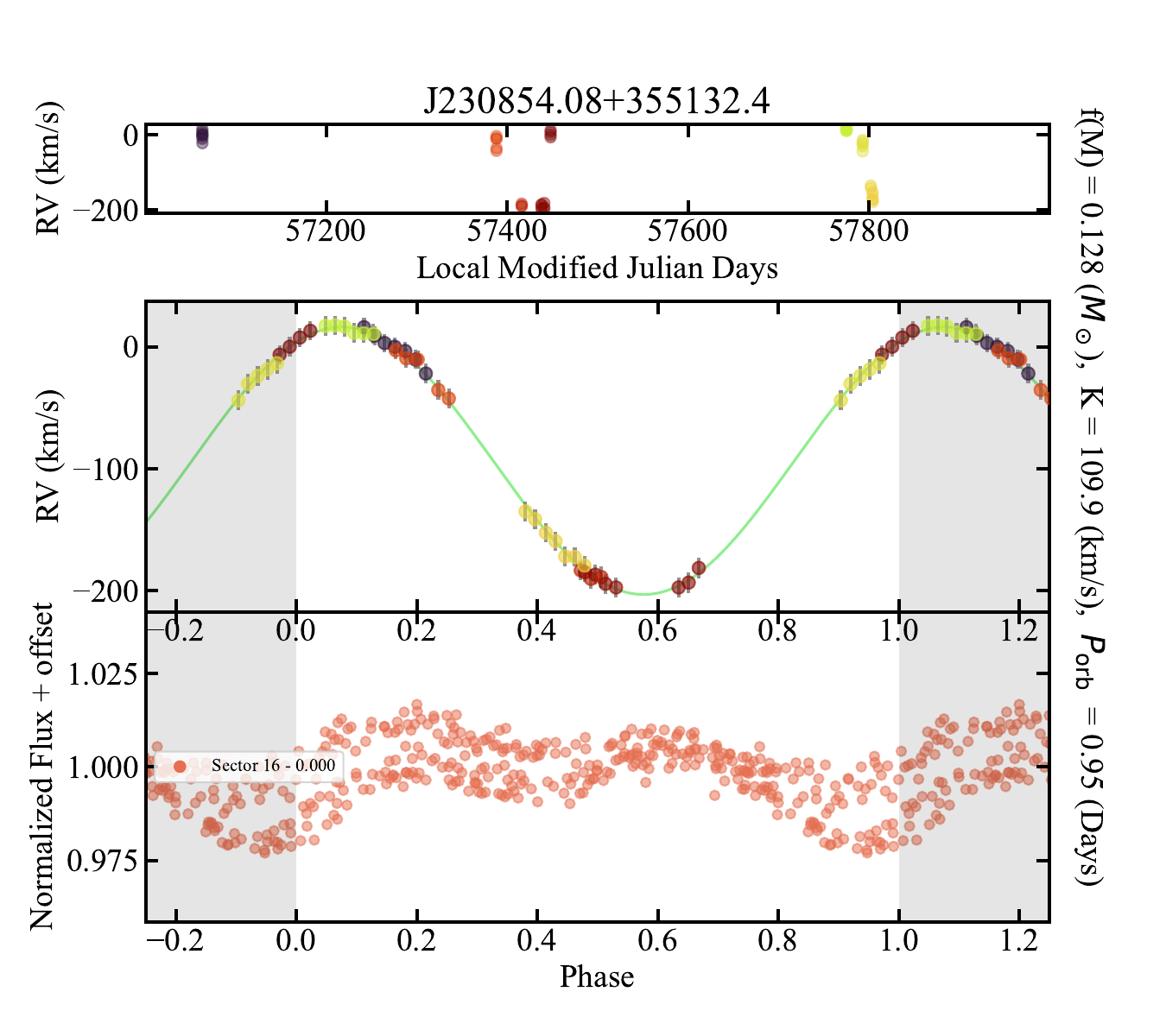}
\caption{Radial velocity curves for target No.3-14.\label{rv3_14}}
\end{figure}

\begin{figure}[h]
\centering
\includegraphics[scale=0.25]{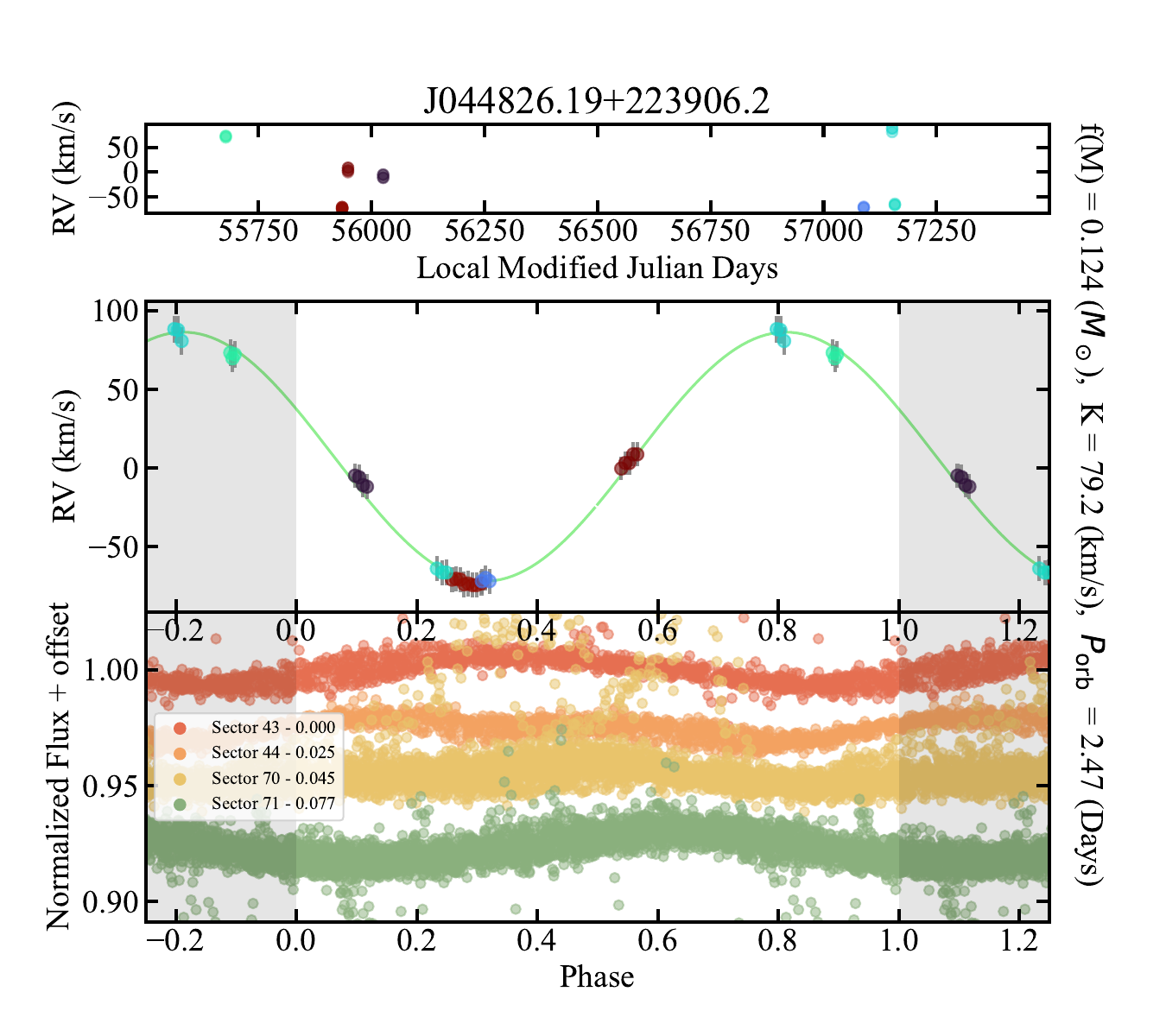}
\includegraphics[scale=0.25]{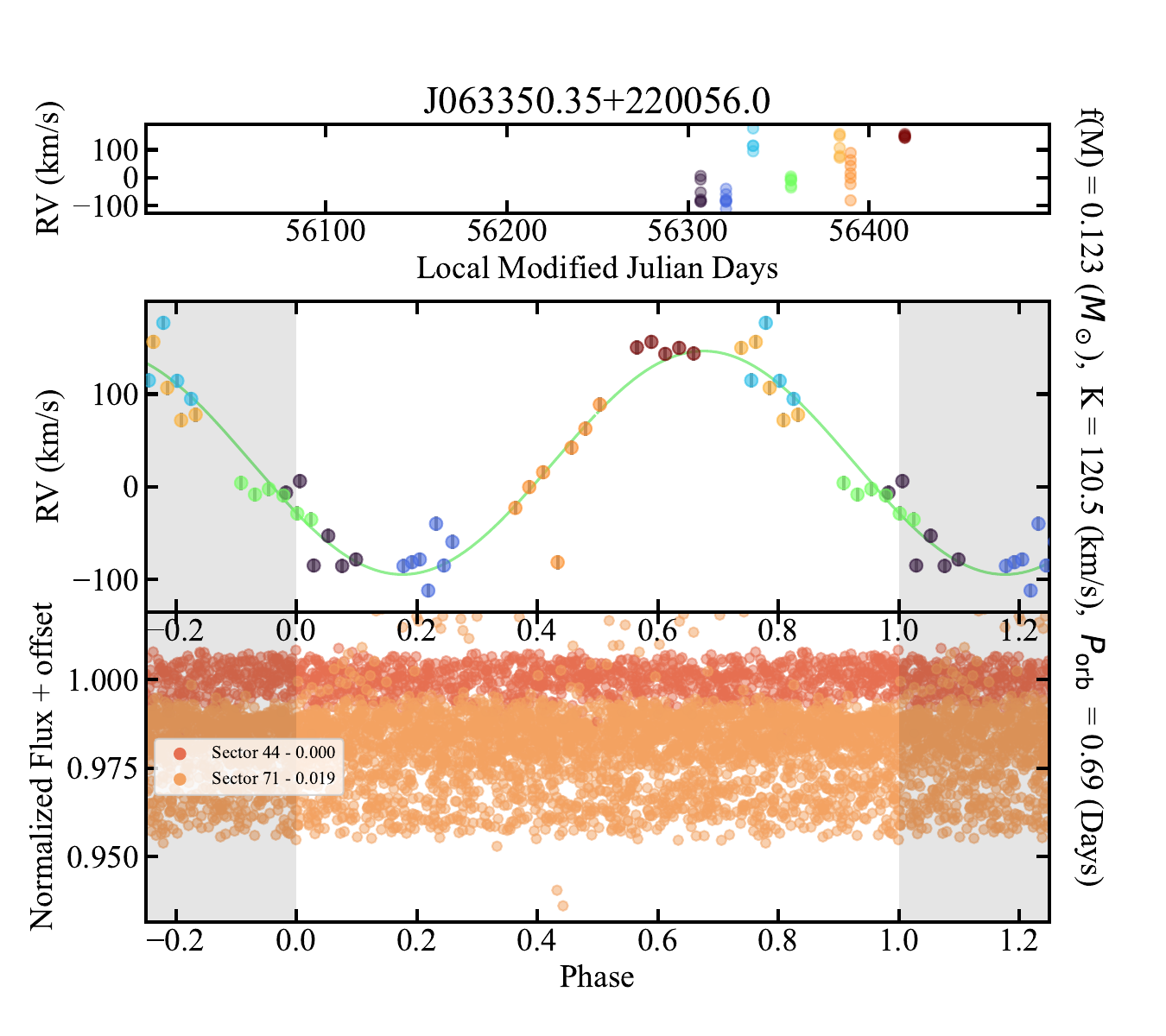}
\includegraphics[scale=0.25]{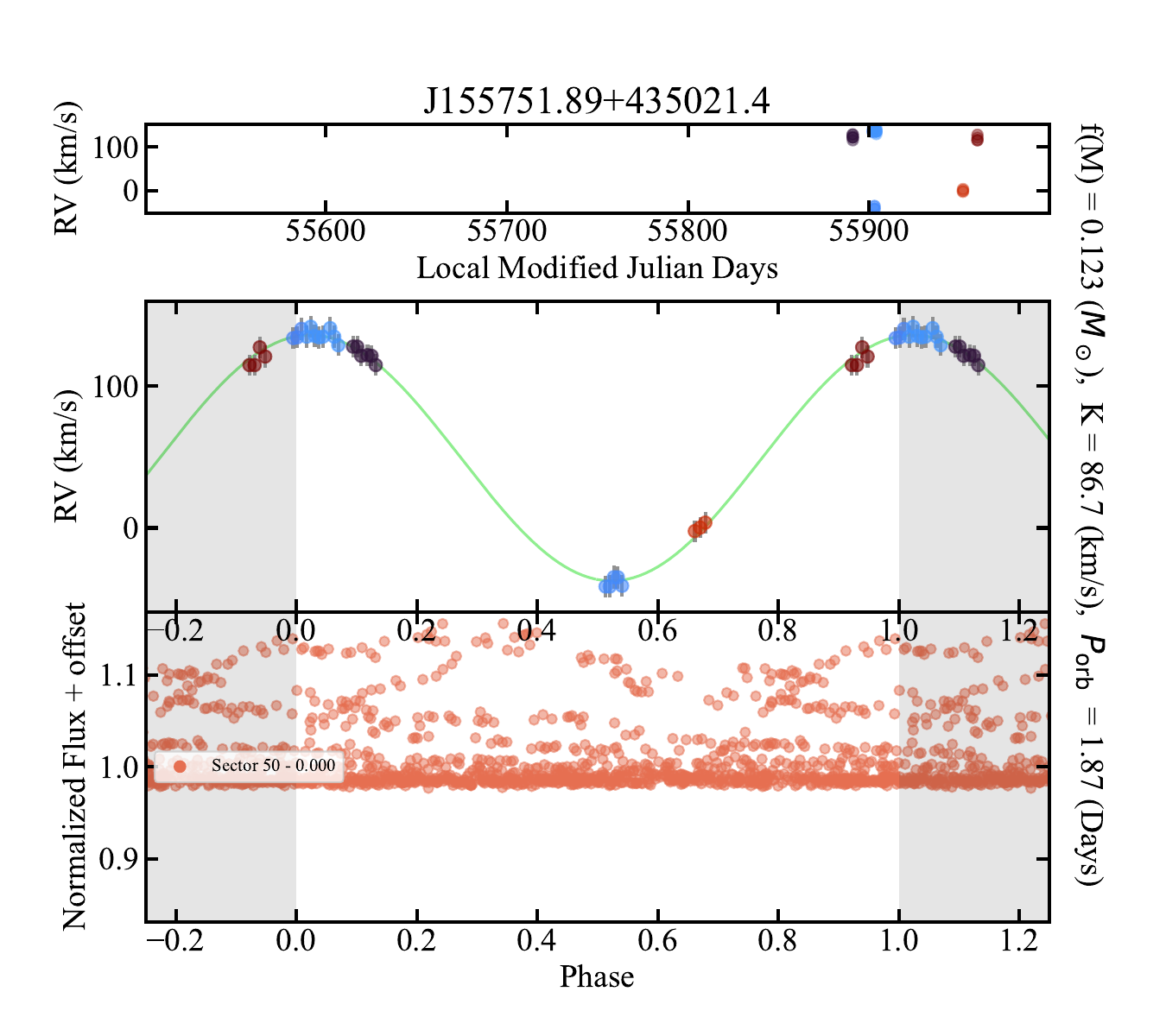}
\includegraphics[scale=0.25]{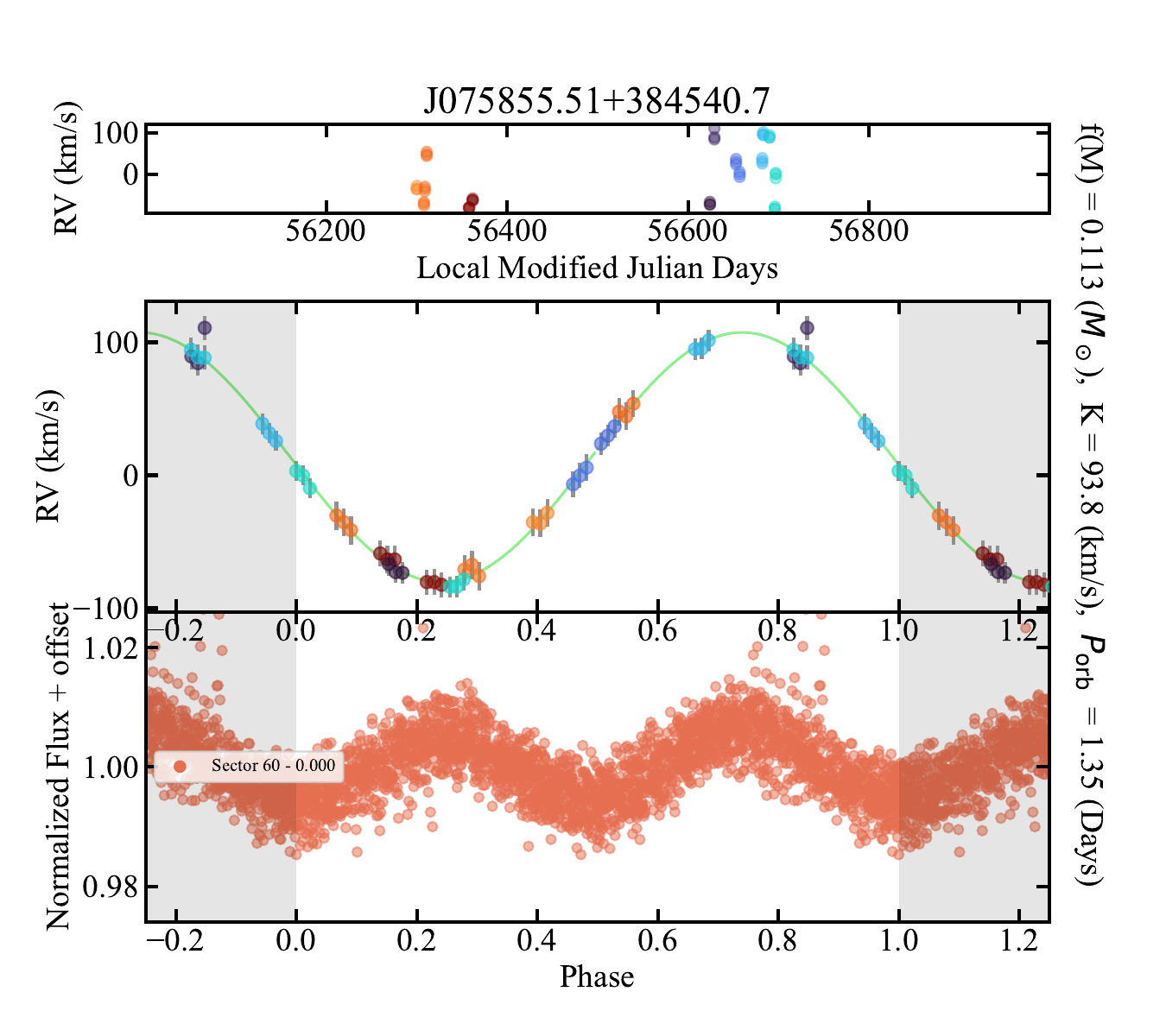}
\includegraphics[scale=0.25]{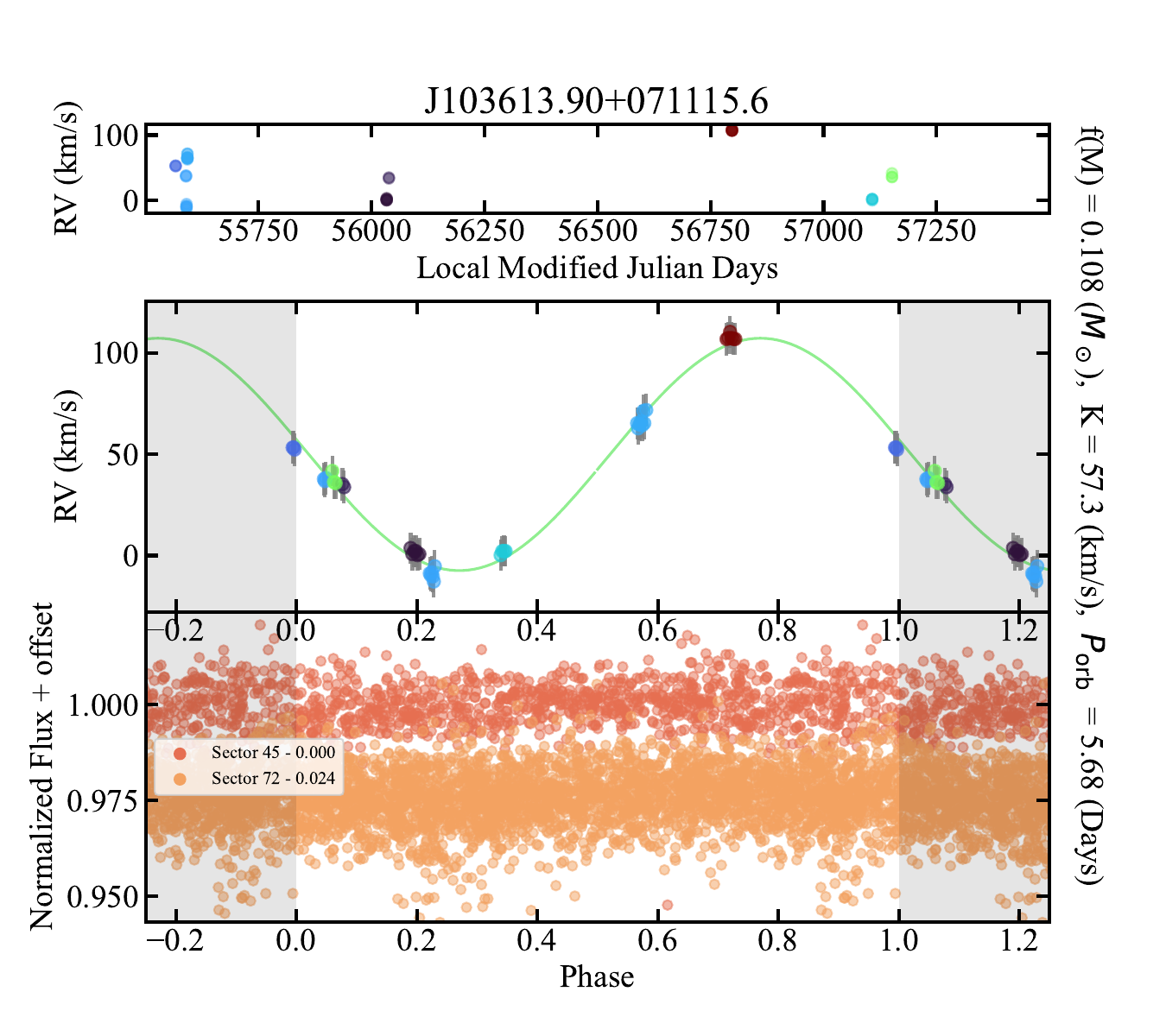}
\includegraphics[scale=0.25]{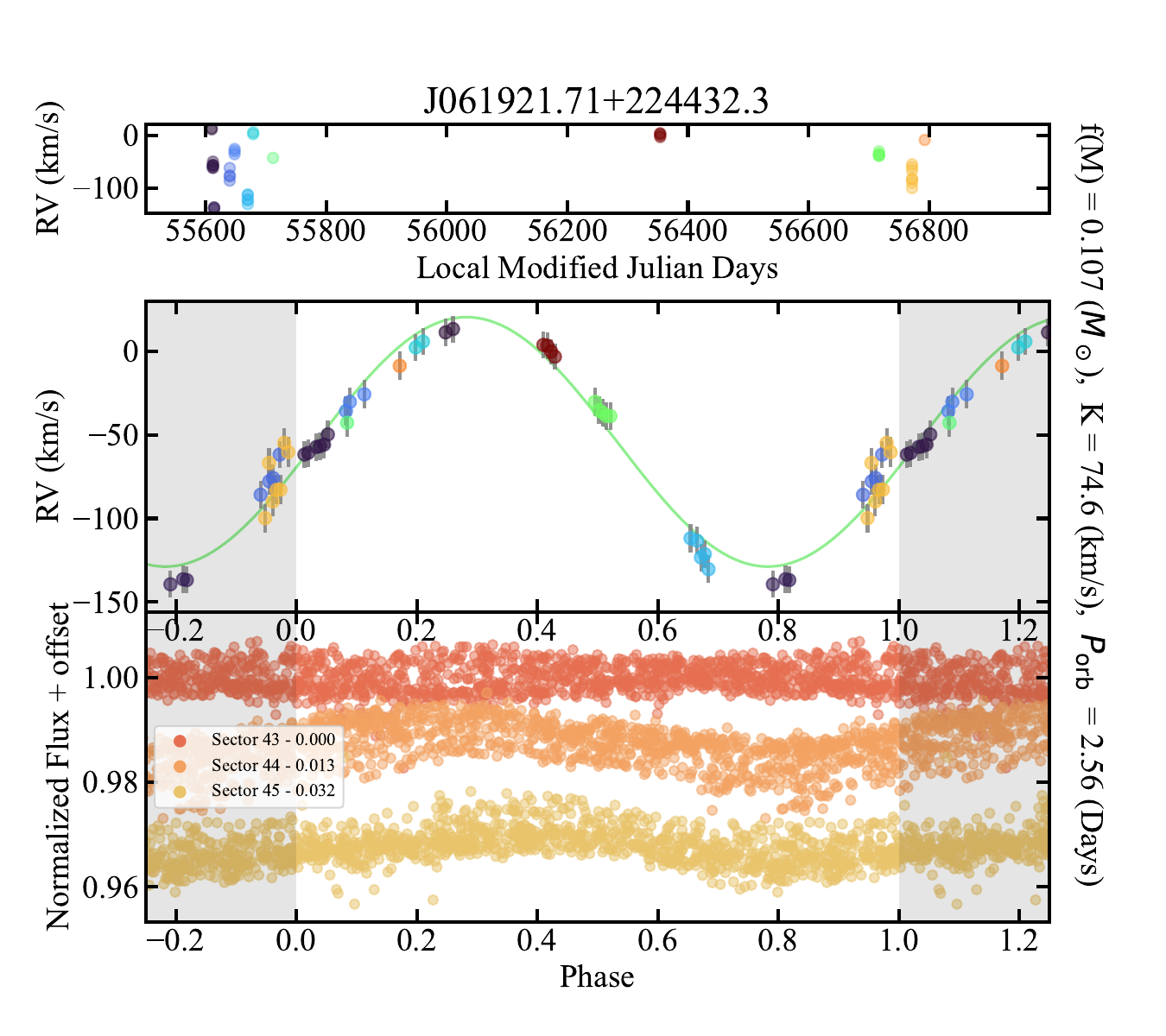}
\includegraphics[scale=0.25]{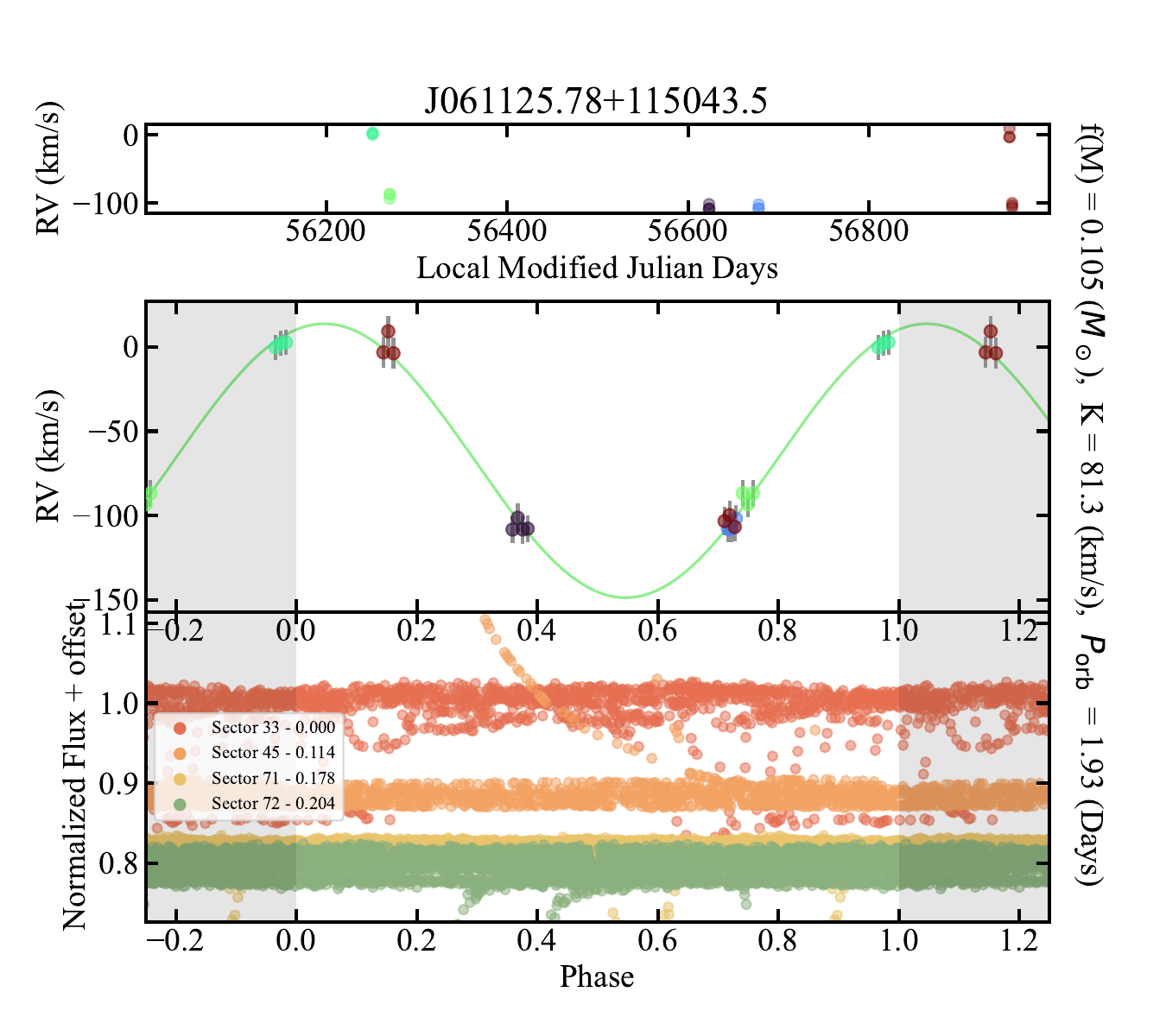}
\includegraphics[scale=0.25]{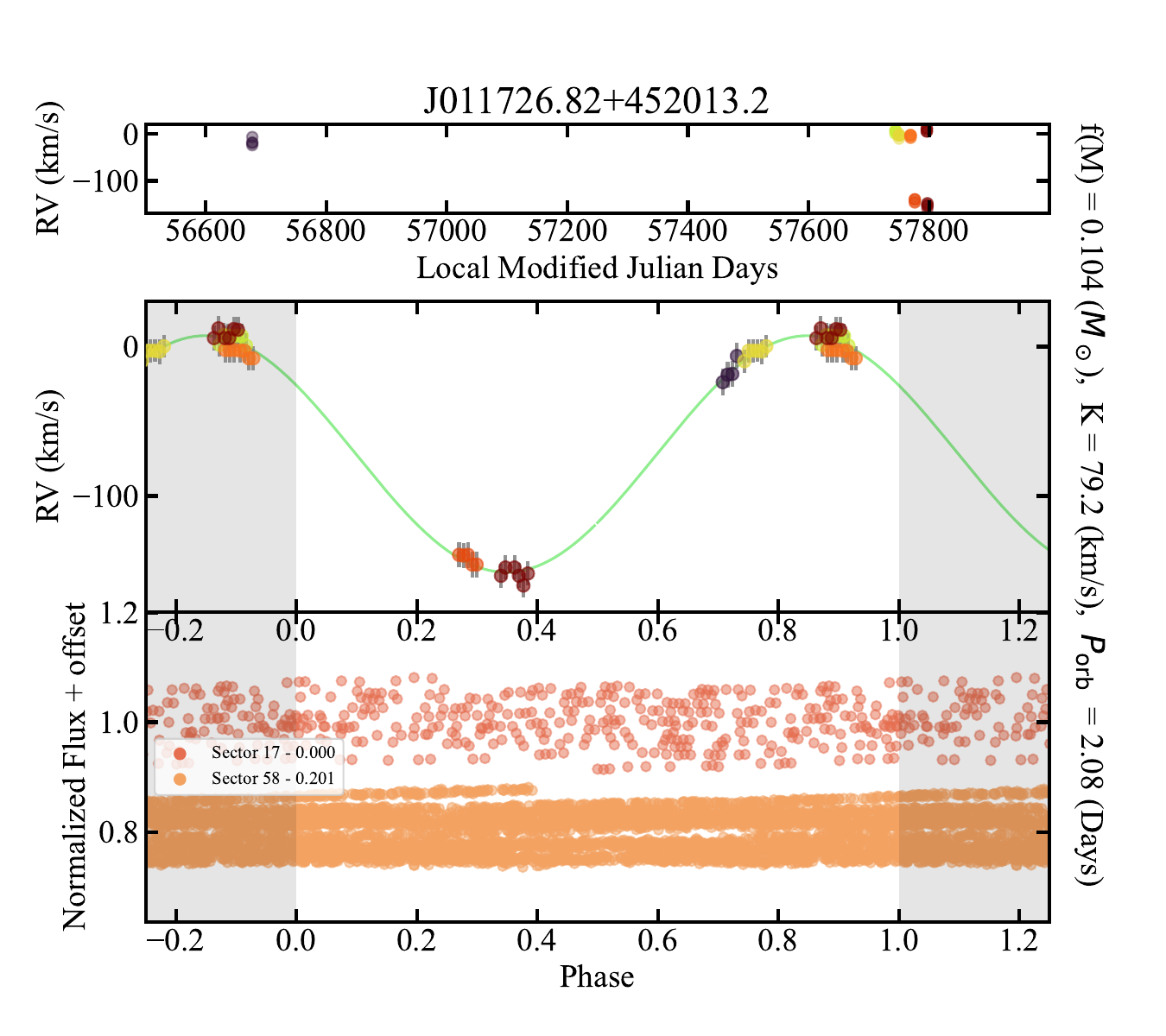}
\includegraphics[scale=0.25]{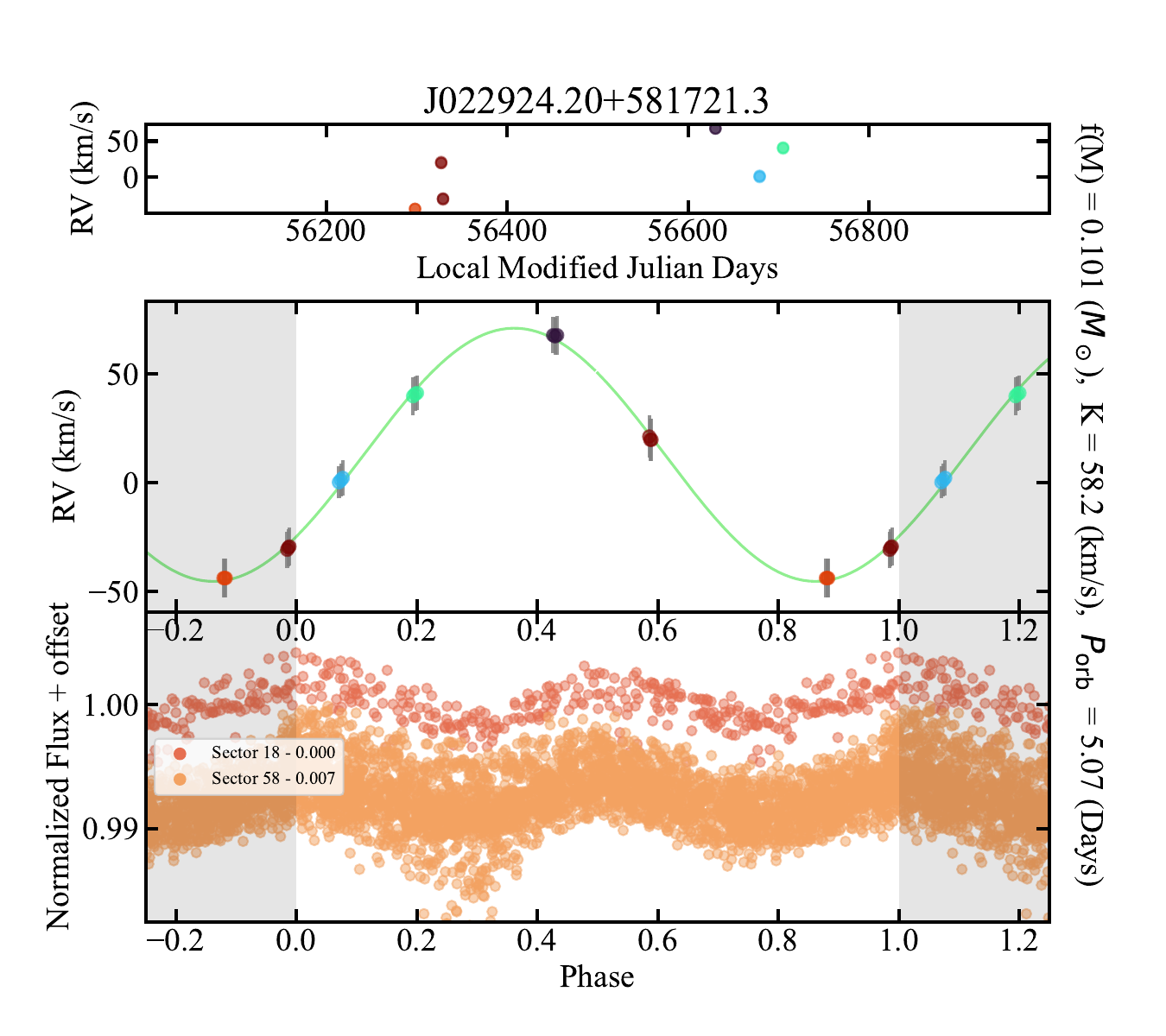}
\includegraphics[scale=0.25]{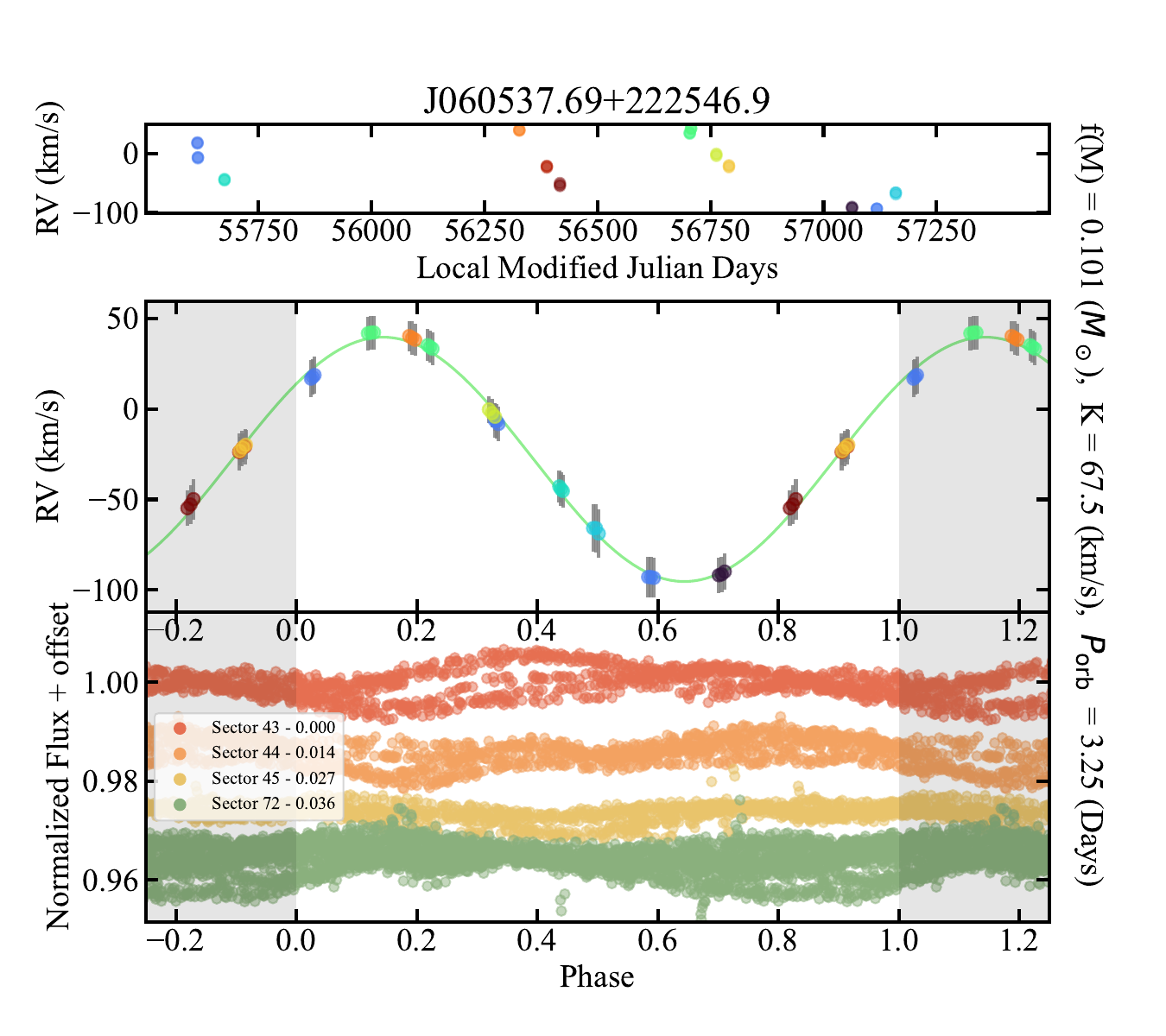}
\includegraphics[scale=0.25]{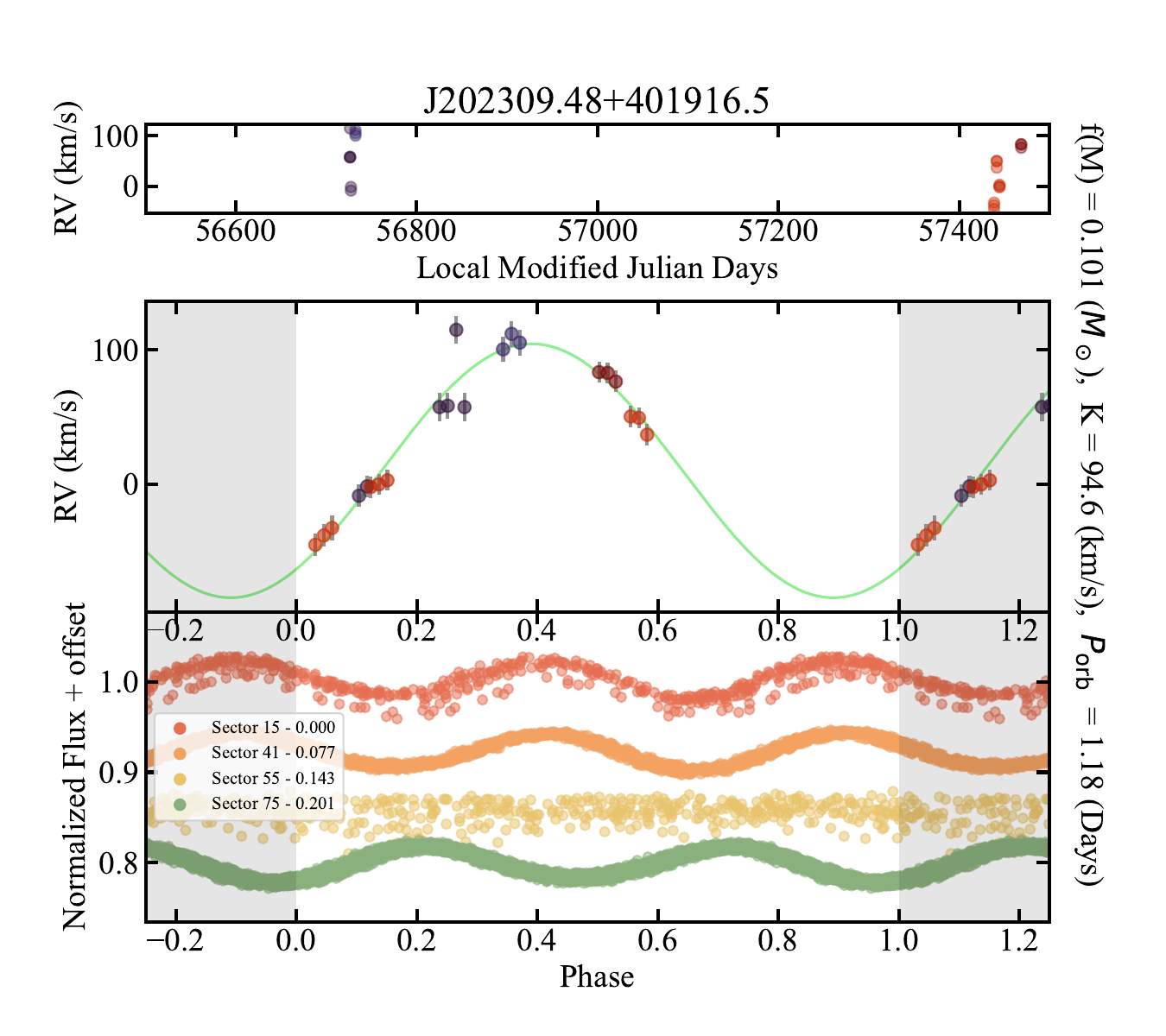}
\caption{Radial velocity curves for target No.15-17 and No.19-26.\label{rv15_26}}
\end{figure}

To assess the reliability of the orbital fits and their uncertainties, here we show radial velocity curves of candidates with mass function larger than 0.1 \msun\, in Figure \ref{rv3_14} and Figure \ref{rv15_26}.


\end{CJK*}
\end{document}